\documentclass[11pt,a4paper]{article}

\usepackage{amsmath}
\usepackage{braket}
\usepackage{epigraph}
\usepackage{jheppub}
\usepackage{tensor}
\usepackage{slashed}
\usepackage{epstopdf}
\usepackage[toc,page]{appendix}
\usepackage{amsmath}
\usepackage{slashed}
\usepackage{amssymb}
\usepackage{braket}
\usepackage{graphicx}
\usepackage{amsmath}
\usepackage{braket}
\usepackage{epigraph}
\usepackage{tensor}
\usepackage{enumerate}
\usepackage{subfig}
\usepackage{verbatim}
\usepackage{pstricks}
\usepackage{axodraw4j}
\usepackage{color}

\date{}

\title{Leptogenesis from loop effects in curved spacetime}
\author{Jamie I. McDonald and Graham M. Shore}
\affiliation{Department of Physics,\\
Swansea University,\\
Swansea, SA2 8PP,  UK.}
\emailAdd{pymcdonald@swansea.ac.uk}
\emailAdd{g.m.shore@swansea.ac.uk}

\notoc

\abstract{We describe a new mechanism -- radiatively-induced gravitational leptogenesis -- for generating the
matter-antimatter asymmetry of the Universe. We show how quantum loop effects in C and CP violating theories 
cause matter and antimatter to propagate differently in the presence of gravity, and prove this is forbidden in 
flat space by CPT and translation symmetry. This generates a curvature-dependent chemical potential for leptons, 
allowing a matter-antimatter asymmetry to be generated in thermal equilibrium in the early Universe. 
The time-dependent dynamics necessary for leptogenesis is provided by the interaction of the virtual self-energy 
cloud of the leptons with the expanding curved spacetime background, which violates the strong equivalence 
principle and allows a distinction between matter and antimatter. We show here how this mechanism is realised
in a particular BSM theory, the see-saw model, where the quantum loops involve the heavy sterile neutrinos
responsible for light neutrino masses. We demonstrate by explicit computation of the relevant two-loop 
Feynman diagrams how these radiative corrections display the necessary dependence on the sterile neutrino 
masses to generate an asymmetry, and show how the induced lepton asymmetry may be sufficiently large to play 
an important r\^ole in determining the baryon-to-photon ratio of the Universe. }

\begin{document}
\maketitle
\newpage
\section{Introduction}\label{YFandRIGL}

Recently, we presented a new mechanism \cite{RIGLletter} for generating matter-antimatter asymmetry in the 
Universe by exposing a deeper connection between matter, antimatter and gravity. Our central finding was that 
at the quantum loop level, matter and antimatter propagate differently in the presence of gravity when C and CP 
are violated --  a phenomenon we showed to be forbidden in flat space by translation invariance and CPT. 
This leads to a difference in the dispersion relations for matter and antimatter, which manifests itself in the form 
of a chemical potential for baryons or leptons in an effective action, and leads to a mechanism for generating the 
baryon asymmetry of the Universe without invoking out-of-equilibrium decays of heavy BSM particles. 
In this paper, we explain these ideas in greater detail and give an extended account of the calculations which 
underlie them. 

This mechanism -- radiatively-induced gravitational (baryo) leptogenesis -- is very general, and relies only on the 
breaking of three symmetries:~C, CP and spacetime translation invariance. This is realised in a cosmological setting 
by the Hubble expansion of the Universe, which provides the time-dependent dynamics necessary to produce a 
matter-antimatter asymmetry.
Consequently, many flat space BSM theories of baryogenesis and leptogenesis, which by virtue of the Sakharov conditions 
\cite{Sakharov} must violate C and CP,  will naturally exhibit our mechanism when minimally coupled to gravity,  
without the need to add new fundamental interactions. Moreover, and remarkably for an intrinsically quantum field
theoretic effect in curved spacetime, it may be strong enough to generate the observed baryon-to-photon ratio of 
the Universe.

In the remainder of this section, we introduce the key ideas of our mechanism in the context of leptogenesis, illustrated
in the popular `see-saw' model \cite{FukYan} in which heavy sterile neutrinos are introduced to provide masses 
for the SM neutrinos. We contrast the way in which leptogenesis arises in this model through the coupling of quantum 
loops involving {\it virtual} sterile neutrinos to the cosmological gravitational field with the conventional mechanism,
where the lepton asymmetry arises from the decays of on-shell sterile neutrinos as they fall out of equilibrium as the 
Universe cools.  

After this overview, we describe our approach in detail from first principles, beginning in section \ref{propagation}, 
where we discuss radiative corrections to matter and antimatter propagation in flat and curved backgrounds. 
Here, we demonstrate that when C and CP are violated, the breaking of translational invariance by gravity leads to a 
difference in the propagation of matter and antimatter. We also provide a simple proof, at the level of both 
S-matrix elements and correlators, that CPT and translation invariance prevent this situation in flat space. 

In section \ref{EFTcurved} we study the effective field theory generated by these propagators. We show that when 
we integrate out the heavy degrees of freedom to construct a low energy effective Lagrangian, distinct matter and antimatter 
propagation generates a C and CP violating operator coupling the lepton current to the derivative of the Ricci scalar. 
In isotropic cosmologies, this operator leads to a chemical potential between leptons and antileptons, generating a 
lepton asymmetry driven by the expansion of the Universe. Since CP violation in the see-saw model arises first at 
fourth order in the complex coupling between the light and sterile neutrinos, the lepton number asymmetry is generated
by two-loop Feynman diagrams contributing to the lepton self-energy. The evaluation of the relevant Feynman 
diagrams is presented in section \ref{Results}, where we show explicitly how the required asymmetry arises.

The implications for leptogenesis in the early Universe are discussed briefly in section \ref{leptogenesis},
where we evaluate the quasi-equilibrium lepton asymmetry in a radiation-dominated FRW universe and 
show that this may be sufficiently large to play an important r\^ole in determining the observed baryon-to-photon
ratio of the Universe.

\subsection{The mechanism}\label{YFandRIGL}

Although our mechanism is very general, for clarity we illustrate it in this paper within a particular model -- 
the see-saw model, first proposed as a means of obtaining a baryon asymmetry 
via leptogenesis by Fukugita and Yanagida \cite{FukYan}. The corresponding minimally coupled Lagrangian is given by
\begin{align}
\mathcal{L} = \mathcal{L}_{EW} +    \sqrt{-g}\left[ \overline{N} i \slashed{D} N+ \lambda_{i \alpha} \bar{\ell}_i\phi  N_\alpha   + 
\frac{1}{2}\overline{\left(N^c\right)} 
\, M \,  N  + \mbox{h.c.}\right]  \ .
\label{FYmodel}
\end{align}  
where $\sqrt{-g}$ is the square root of the metric determinant, and $D$ is the spinor covariant derivative.  
In this model, $\ell_i$ ($i = e, \mu , \tau$) are the light, left-handed lepton doublets and $\phi$ is the Higgs 
field\footnote{In this notation, the Higgs doublet $\tilde{\phi}$ appearing in the SM lepton sector is related by 
$\phi^a = \epsilon^{ab} \tilde{\phi}^{\dagger b}$.} which couples to heavy right-handed sterile neutrinos $N_\alpha$ 
with non-degenerate masses $M_\alpha$ ($\alpha=1,\ldots n$). Crucially, the Yukawa couplings $\lambda_{i\alpha}$ 
contain irremovable complex phases, providing a source of C and CP violation.  The first Sakharov condition is realised 
in two parts:~the Yukawa interaction violates lepton number by one unit, allowing the creation of matter-antimatter 
asymmetry in the lepton sector. This is then converted to a baryon asymmetry of the same magnitude via sphaleron 
processes \cite{Klinkhamer:1984di, Kuzmin:1985mm}, which are in equilibrium at high temperatures in the early Universe. 

We note also at this point that that the Majorana mass term for the heavy neutrinos gives two classes of propagators, 
\textit{charge-violating} propagators $S^\times_{\! \alpha}(x,x')  = \braket{N (x)\overline{N^c}(x')}$ and 
\textit{charge-conserving} propagators $S_{\! \alpha}(x,x')=\braket{N_\alpha (x)\overline{N}_\alpha(x')}$, where the $C$ 
script denotes the Dirac charge conjugate. All diagrams can be expressed in terms of this basis of sterile neutrino 
propagators. In flat space, translation invariance allows us to write them in momentum space as
\begin{equation}\label{flatprops}
S_\alpha(p) = \frac{i \slashed{p}}{p^2 - M_\alpha^2}, \qquad
S^\times_\alpha(p) = \frac{i M_\alpha}{p^2 - M_\alpha^2} \ .
\end{equation} 

We now describe how the third Sakharov condition, namely a departure from equilibrium, is satisfied 
in the traditional sterile neutrino decay scenario, and then how leptogenesis is realised in our own mechanism.

\subsection*{Leptogenesis from heavy neutrino decays}

The traditional leptogenesis model neglects gravitational effects in the Lagrangian, with all relevant amplitudes calculated 
in flat space. In the simplest leptogenesis scenario \cite{FukYan, KolbTurner, Buchmuller, Pedestrians}, 
one assumes a thermal initial abundance of sterile
neutrinos\footnote{There are other scenarios \cite{Pedestrians} 
for the initial state of the sterile neutrinos, but in all cases, the
lepton asymmetry is generated when they depart 
from their equilibrium distribution.  }. At early times, when the temperature is sufficiently 
high ($T\gtrsim M_1$), decays $N_1 \rightarrow \ell \phi$ and inverse decays $\phi \ell \rightarrow N_1$ are in equilibrium 
and the sterile neutrinos remain thermalised with relativistic number densities. However, as the temperature drops
to $T \lesssim  M_1$, the inverse decays are Boltzmann suppressed by a factor $e^{-M_1/T}$ and become inefficient. 
For a short time, this leaves the sterile neutrinos slightly over-abundant compared to their non-relativistic equilibrium 
distribution, until eventually they decay. This is the out-of-equilibrium process necessary to satisfy the Sakharov condition 
and allow leptogenesis.

\begin{figure}[h!]
\centering
\includegraphics[scale=0.45]{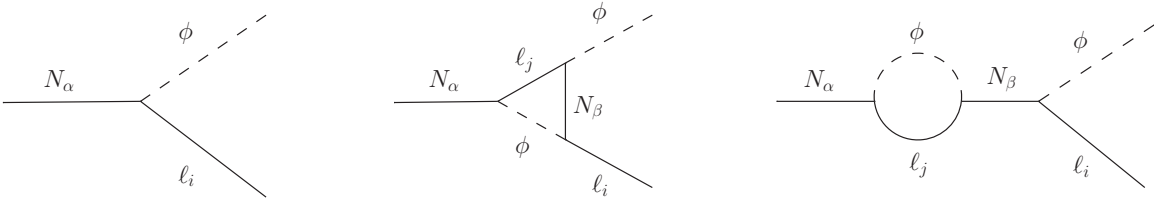}
\caption{Flat space diagrams contributing to out-of-equilibrium
  decays. The second and third diagrams 
  give $f(x)$ and $g(x)$ respectively.}
\label{decays}
\end{figure}
An interference between the tree-level and one-loop decay amplitudes shown in figure \ref{decays} gives a difference 
in the production rates of leptons and antileptons at the time of the heavy decays, as characterised by the well-known 
quantity \cite{Buchmuller}
\begin{equation}\label{decaysYF}
\epsilon_\alpha  = - \frac{1}{8\pi} \sum_{\beta \neq \alpha}
\frac{\mbox{Im}[ (\lambda^\dagger \lambda)^2_{\alpha
    \beta}]}{(\lambda^\dagger 
\lambda)_{\alpha \alpha}}  \left[ f\left(\frac{M_\beta^2}{M_\alpha^2}\right) + g\left(\frac{M_\beta^2}{M_\alpha^2}\right) \right],
\end{equation}
where
\begin{align}
f(x) &= \sqrt{x}\left( 1  -(1+x) \ln \left(\frac{1+x}{x} \right)\right), \nonumber \\
g(x) &= \frac{\sqrt{x}}{1-x}.
\end{align}
 In this way, a lepton asymmetry is generated at the point when the sterile neutrinos fall out of equilibrium.

\subsection*{Leptogenesis from radiatively-induced gravitational couplings}

\begin{figure}[h!]
\centering
\includegraphics[scale=0.45]{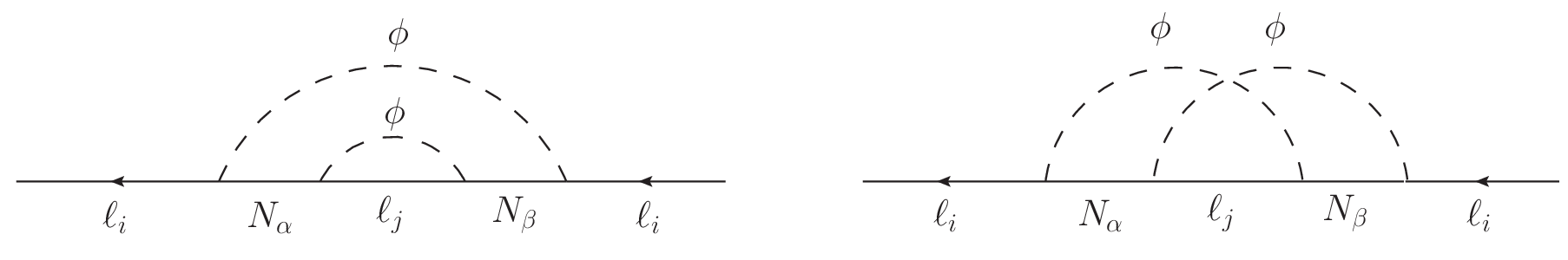}
\caption{Curved space two-loop self-energies which contribute to radiatively-induced gravitational leptogenesis.}
\label{2loop}
\end{figure}

In our scenario, the time-dependent dynamics necessary to distinguish the dispersion relations for leptons
and antileptons and induce the matter-antimatter asymmetry arises from the expansion of the Universe itself,
through the gravitational coupling to the virtual self-energy cloud screening the leptons at the quantum 
loop level. Here, the heavy sterile neutrinos play an altogether different role. Rather than generating the 
lepton number asymmetry through their out-of-equilibrium decays, they contribute to our mechanism only as
 virtual particles mediating the propagation of the light leptons, as shown by the diagrams in 
figure \ref{2loop}. The propagators in these self-energy diagrams are the appropriate curved space Green functions
(see, {\it e.g.}, ref.~\cite{RIGLletter}) derived from the minimally-coupled Lagrangian (\ref{FYmodel}).
Crucially, the introduction of a scale associated with the mass of the virtual particles in the loops
allows a direct coupling to the curvature, which violates the strong equivalence principle and 
allows the leptons and antileptons to propagate differently.

When the heavy sterile neutrinos are integrated out from the Lagrangian (\ref{FYmodel}), these 
self-energy diagrams produce the following C and CP violating operator in the low-energy effective action:
\begin{equation}\label{Operator}
\mathcal{L}_i = \partial_\mu R \, \left( \bar{\ell}_i \gamma^\mu \ell_i \right)\,
\sum_{\alpha,  \,  \beta, \,   j} \frac{ \mbox{Im}\left[
    \lambda^\dagger_{\beta i} \lambda_{i\alpha} \lambda^\dagger_{\beta
      j} \lambda_{j \alpha} \right]}{3 M_\alpha M_\beta} I_{[\alpha \beta] },
\end{equation}
where the loop factor $I_{\alpha \beta} = I(M_\alpha,M_\beta)$, which we shall discuss at length in subsequent sections, 
is a function of the two sterile neutrino masses $M_\alpha$ and $M_\beta$ in the loop. 
In a homogeneous and isotropic universe, the spatial derivatives of $R$ vanish, leading to a chemical potential 
for each lepton generation of the form
\begin{equation}\label{mui}
\mu_i  = \dot{R} \sum_{\alpha,  \,  \beta, \,   j} \frac{  \mbox{Im}\left[
    \lambda^\dagger_{\beta i} \lambda_{i\alpha} \lambda^\dagger_{\beta
      j} \lambda_{j \alpha} \right]}{3 M_\alpha M_\beta} I_{[\alpha \beta] }.
\end{equation}
This changes the equilibrium distributions of leptons and antileptons and,
after summing over lepton generations, we get a total lepton asymmetry\footnote{
See, for example, \cite{Kolb:1979qa} appendix C for a compendium of useful formulae
for number and energy densities for particles of different statistics in thermal equilibrium.}
\begin{equation}\label{RIGL}
n_L = \frac{T^2 g_{\ell}}{6}\dot{R} \sum_{\alpha ,\beta} \frac{  \mbox{Im}[ (\lambda^\dagger 
\lambda)^2_{\alpha \beta}]}{3 M_\alpha M_\beta}  I_{\alpha \beta}.
\end{equation}

Previously, C and CP violating operators of this kind have been added by hand 
\cite{Cohen, Davoudiasl:2004gf, Lambiase:2006md,  Lambiase:2011by,  Lambiase:2013haa, Lambiase:2015}, 
with the assumption that they may arise from a complete theory of gravity. 
However, without an obvious source of C and CP violation in the underlying theory, it remains unclear how these 
operators would actually arise in an effective theory of quantum gravity, and with what magnitude.
Instead, we demonstrate here how they are generated in a simple and elegant fashion directly from loop
effects in a BSM quantum field theory in curved spacetime.
Furthermore, we have a readily accountable source of CP violation from the Yukawa couplings. 

\begin{figure}[h]
\centering
\includegraphics[scale=0.3]{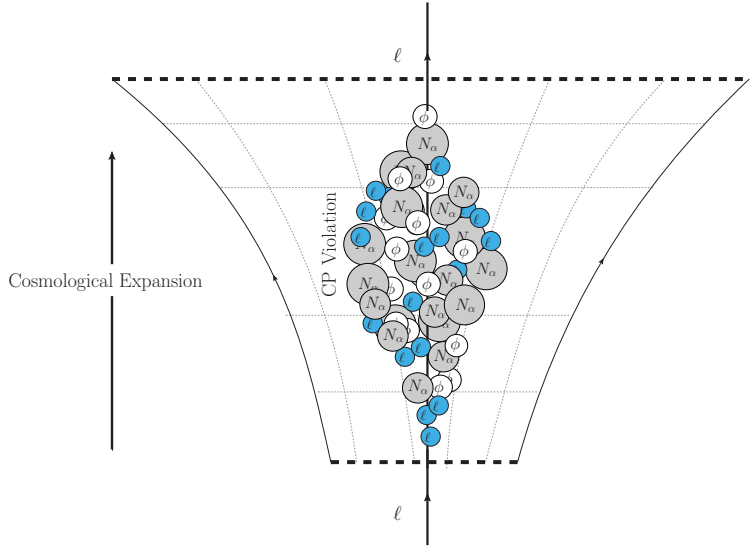}
\caption{A schematic drawing illustrating how the lepton becomes sensitive both to CP violation 
and to the time-dependent nature of the background at the loop level. }
\label{cloud}
\end{figure}

We will describe this effect in great detail in the remainder of this paper, but for now, 
we emphasise three crucial radiatively-induced effects which allow the realisation of our mechanism 
and the generation of the operator (\ref{Operator}).
\begin{enumerate}
\item As in the traditional scenario, it is only at the loop level that one becomes sensitive to the 
complex phases of the Yukawa couplings, so that C and CP violating operators such as (\ref{Operator}) 
are only produced above some minimum number of loops needed to expose the CP violation.

\item The operator (\ref{Operator}) manifestly violates the strong equivalence principle (SEP), 
allowing a distinction between the gravitational effect on different particles.
Minimal coupling of the Lagrangian ensures that at tree level, the SEP still holds. 
However, at the loop level, the insertion of curved space propagators in figure \ref{2loop} means that 
virtual particles probe the details of the background, causing the leptons to become sensitive to curvature effects. 
As a result, even in an inertial frame, radiative effects force the particle to become sensitive to curvature, 
permitting the existence of SEP violating operators (\ref{Operator}). The interpretation of this effect 
is that the screening cloud causes the lepton to acquire an effective size and experience tidal forces, 
realised by couplings of various curvature tensors to the lepton fields in the effective action, such as (\ref{Operator}).
Together with C and CP violation, this SEP-violating effect also distinguishes between the leptons
and antileptons.

\item In one form or another, time dependence of the dynamics is a necessary ingredient in any 
mechanism for producing a matter-antimatter asymmetry. In the sterile neutrino decay mechanism, 
this is realised according to the conventional Sakharov condition by a non-equilibrium process. 
In contrast, gravitational leptogenesis introduces time dependence through the coupling to an
evolving background gravitational field \cite{Cohen}. The novel feature of our mechanism is that this
sensitivity to the background arises dynamically, through explicitly quantum field theoretic effects 
occurring naturally at loop level. This elucidates why the operator (\ref{Operator}) responsible for the 
matter-antimatter asymmetry depends on the non-vanishing of the time derivative of the curvature. 
In this sense, the leptons inherit the time-dependent dynamics of the background. This is illustrated 
schematically in figure \ref{cloud}.
\end{enumerate}

Having highlighted some of the key similarities and differences between the traditional mechanism and 
our new approach, as embodied by the formulae (\ref{decaysYF})   and (\ref{RIGL}) respectively, 
we spend the remainder of this paper examining in greater detail the origin of our effect, beginning 
with a discussion of propagation.

\newpage
\section{Matter-antimatter propagation in flat and curved spacetime}\label{propagation}

We begin with the following motivation, namely that the operator (\ref{Operator}) causes a splitting in the 
spectra of matter and antimatter. Consider a quasi-plane wave solution for the lepton field $\ell(x) = u(x) e^{-i \Theta(x)}$, 
so that the momentum of the particle is given by $p_\mu = \partial_\mu \Theta$. As discussed in \cite{McDonald:2014}, 
when an operator of the form $b \, \partial_\mu R j^\mu$, coupling the derivative of the Ricci scalar
to the lepton current, is added to the Dirac action, 
it gives different dispersion relations for particles and antiparticles.
To see this, note that the modified equations of motion
\begin{equation}
i \slashed{D} \ell  + b \, \partial_\mu R \gamma^\mu \ell =0, \qquad  \quad i\slashed{D} \ell^c  - b \, 
\partial_\mu R \gamma^\mu \ell^c=0
\end{equation}
are solved in the context of the eikonal approximation \cite{McDonald:2014} by 
\begin{equation}
\ell \sim u(x) e^{- i (\Theta -b R)}, \qquad  \ell^c \sim v(x) e^{- i (\Theta +b R)} 
\end{equation}
leading to the dispersion relations 
\begin{equation}
\left(  p_\mu \pm b\, \partial_\mu R \right)^2 =0, 
\end{equation}
for the leptons and antileptons respectively.
This causes a difference of energies between matter and antimatter -- a picture which is consistent with the
 interpretation (\ref{mui}) of (\ref{Operator}) as a chemical potential, which also corresponds to an energy-cost 
difference between particles and antiparticles. 

We see then that the existence of this operator in an effective action necessarily implies that matter and antimatter 
propagate differently through the gravitational medium. This motivates the first step in the description of our mechanism, 
which involves a study of the propagation of matter and antimatter in flat and curved backgrounds.

\subsection{Propagation in translation invariant backgrounds}\label{flatpropagation}

In this section, we show, in two different ways, that matter and antimatter must propagate identically in translation invariant 
backgrounds. First, consider the transition amplitude for a particle to propagate between $x$ and $x'$:
\begin{equation}
f_s(x',x) = \braket{\psi(x') ,s  | \psi(x) , s }.
\end{equation} 
Here, $\psi(x')$ and $\psi(x)$ denote a particle at $x$ and $x'$, and  $s$ labels helicity (spin) for massless (massive) particles. 
The corresponding amplitude for antiparticles is
\begin{equation}\label{AP}
f_s^c(x',x) = \braket{\psi^c(x') ,s  | \psi^c(x) , s },
\end{equation} 
where the $C$ superscript denotes charge conjugate states. CPT symmetry is realised  by an anti-unitary operator 
$\Theta$ in such a way that the inner product, represented by the bracket notation, is preserved under the action of 
$\Theta$ on each argument, together with an overall complex conjugation \cite{Streater}, \textit{i.e.},
\begin{equation}
 \braket{\psi(x') ,s  | \psi(x) , s } = \left\{ \left( \bra{\psi(x') , s  } \Theta \right) \left(  \Theta \ket{  \psi(x) , s } \right) \right\}^*
\end{equation}
Since $s$ is odd under CPT, we have $\Theta \ket{\psi(x),s} = \ket{\psi^c(-x),-s}$, and so, after complex conjugating, we get
\begin{equation}\label{xx'}
 \braket{\psi(x') ,s  | \psi(x) , s } = \braket{\psi^c(-x) , -s  |  \psi^c(-x') , -s }\equiv f^c_{-s} (-x,-x'),
\end{equation}
where in the rhs, we used the definition (\ref{AP}). Finally, we invoke translation symmetry which implies 
$f^c_{-s} (x',x) = f^c_{-s} (x' - x)$, \textit{i.e.}, the transition amplitudes are functions of the relative position of the two points. 
This means that $f^c_{-s}(x',x) = f^c_{-s}(-x,-x')$ and hence, from (\ref{xx'}), that
\begin{equation}\label{fisfc}
f_s(x',x) = f_{-s}^c(x',x),
\end{equation}
establishing that the transition amplitude for a particle with spin/helicity $s$ to go from $x$ to $x'$, is the same 
as the transition amplitude for an antiparticle with spin/helicity $-s$ to go from $x$ to $x'$. 

This is precisely the relevant statement for leptogenesis, since neutrinos (antineutrinos) have positive (negative) helicity, 
and shows that neutrinos and antineutrinos propagate identically in translation invariant backgrounds. Similarly,
 for massive particles, this would mean that spin up (down) particles propagate the same as spin down 
(up) antiparticles respectively, so that, averaging over spins, there is no difference in propagation.  
Notice the result (\ref{fisfc}) is non-perturbative and holds generically in any theory satisfying translation invariance. 

We now demonstrate this result explicitly in the see-saw model at the correlator level by studying the lepton 
and antilepton propagators (including radiative corrections) given by
\begin{equation}\label{SSc}
S_{ab}(x',x') = \braket{\ell_a(x') \bar{\ell}_b(x)}, \qquad S^c_{ab}(x',x) = \braket{\ell_a^c(x') \overline{\ell_b^c}(x)},
\end{equation}
where the charge conjugate is given by $\ell = C (\bar{\ell})^T$ and $\overline{\ell^c} = - \ell^T C^{-1} $, 
and the matrix C satisfies $C \left( \gamma^\mu \right)^TC^{-1} =- \gamma^\mu$. As before, 
remembering that the Dirac bra-ket notion represents an inner product, the action of CPT on the propagator is
\begin{align}
S_{ab}(x',x) & =\braket{ \ell_a(x') \bar{\ell}_b (x)} \nonumber \\
& = \braket{ \left(\Theta \ell_a(x') \Theta^{-1} \right) \left( \Theta
    \bar{\ell}_b (x) \Theta^{-1}\right)}^* , 
\end{align}
where
\begin{equation}
\Theta \ell (x') \Theta^{-1} = \gamma^0\gamma_5 C^{-1} \ell^c(-x) \qquad \Theta \bar{\ell}(x') \Theta^{-1} 
=\overline{\ell^c} (-x')C\gamma_5 \gamma^0.
\end{equation}
After performing the overall complex conjugation and, with a little algebra, we arrive at
\begin{equation}
S (x',x) = \gamma_5 C [S^c(-x,-x')]^T C^{-1}\gamma_5 \ .
\end{equation}
As before, translation symmetry allows us to write $S^c(x',x) = S^c(x'-x)$ and permits a momentum space representation
\begin{equation}\label{mom}
S^c(x',x) = \int \frac{d^np}{(2\pi)^n} S^c(p) ^{- i p \cdot (x'-x)}.
\end{equation}
Furthermore, by Lorentz symmetry we must have
\begin{equation}\label{lorentz}
S^c(p) = A(p^2)
  \slashed{p}  + B(p^2),
\end{equation}
for some functions $A$ and $B$, which in general will depend also on $\lambda$, $M_\alpha$ and $M_\beta$ for the 
full propagator. Inserting (\ref{lorentz}) and (\ref{mom}) into (\ref{SSc}), we find, after a little matrix manipulation, that
\begin{equation}\label{SScidentical}
S(x',x) = S^c(x',x).
\end{equation}
This reproduces the same result  (\ref{fisfc}) at the level of correlators, showing that translational invariance forbids 
a difference in lepton and antilepton propagators. Notice that at the correlator level, spin seems not to enter the proof.
This is because, although spin is exchanged under CPT, the field operator is a superposition over spin states, so that 
the flip in spin becomes absorbed into this sum.  

As a consequence of the above results, there can be no asymmetric propagation of matter and antimatter in flat space. 
Conversely, when we relax the constraint of translation invariance, as happens in general gravitational 
backgrounds, we should expect to see a difference in the propagation of matter and antimatter.

\subsection{Propagation in curved backgrounds}

As discussed in section \ref{YFandRIGL}, loop effects lead to a violation of the SEP, which forces lepton propagators to 
become sensitive to a breaking of translational invariance by the background. Of course, to have a difference in the propagation
of leptons and antileptons, it is necessary also to break C and CP violation, which is achieved via the complex phases 
in $\lambda_{i\alpha}$. We now demonstrate explicitly that when these three symmetries are broken, there is a difference 
in the propagation of matter and antimatter. We study this in terms of the self-energies $\Sigma(x',x)$ and $\Sigma^c(x',x)$ 
associated to the lepton and antilepton propagators of (\ref{SSc}).  

\begin{figure}[h!]
\centering
\includegraphics[scale=0.38]{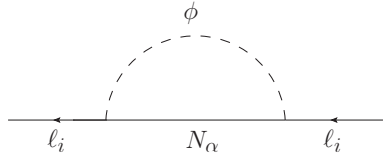}
\caption{One-loop lepton self-energy}
\label{oneloop}
\end{figure}
At one loop (see figure \ref{oneloop}), we immediately see $\Sigma(x',x) - \Sigma^c(x',x)=0$, with
\begin{equation}
\Sigma_i(x',x) = \Sigma^c_i(x',x) =\sum_\alpha\lambda^\dagger_{\alpha i} \lambda^{\!}_{i\alpha}
G(x',x)S_\alpha(x',x) \ ,
\end{equation}
since the couplings occur only in the combination $(\lambda^\dagger \lambda)_{ij}$ for which the diagonal
elements are manifestly real.

\begin{figure}[h!]
\centering
\includegraphics[scale=0.5]{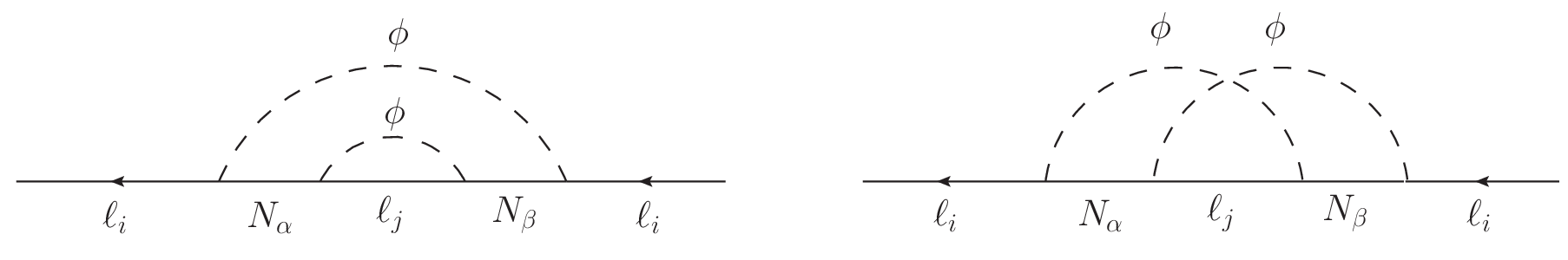}
\caption{Two loop diagrams in curved space which give $\Sigma(x',x)-\Sigma^c(x',x) \neq 0$. }
\label{2loopagain}
\end{figure}
However, at two loops, there are two diagrams (figure \ref{2loopagain}) which give non-zero contributions to 
$\Sigma(x',x)-\Sigma^c(x',x)$. 
For instance, in the case of the charge-violating heavy neutrino
propagators (see section \ref{YFandRIGL}), 
the diagram on the left gives
\begin{align}\label{propagator}
&\Sigma_i(x',x) - \Sigma^c_i(x',x) \nonumber \\
& =   \sum_{\alpha, \, \beta, \, j} 2i  \mbox{Im}
\left[ \lambda^\dagger_{\beta i} \lambda_{i\alpha} \lambda^\dagger_{\beta j} \lambda_{j \alpha} \right]  
G(x',x) \int d^d y \int d^d z \, \, G(y,z)
S^{\times}_{[\alpha} (x',y)S_j(y,z) S^{\times}_{\beta]}(z,x) \ ,
\end{align}
while the one on the right yields
\begin{align}\label{vertex}
&\Sigma_i(x',x) - \Sigma^c_i(x',x) \nonumber \\
&=     \sum_{\alpha, \, \beta, \, j}  2i \mbox{Im}
\left[ \lambda^\dagger_{\beta i} \lambda_{i\alpha} \lambda^\dagger_{\beta j}\lambda_{j \alpha} \right]  
\int d^d y \int d^d z  \,  G(y,x') G(x,z)  S^{\times}_{[\alpha
}(x',y)S_j(y,z) S^{\times}_{\beta]}(z,x) \ .
\end{align}
Notice that we have antisymmetrised over $\alpha$ and $\beta$ in the
integral since  $\mbox{Im}\left[ \lambda^\dagger_{\beta i} \lambda_{i\alpha} \lambda^\dagger_{\beta j} \lambda_{j \alpha} \right] $ 
is antisymmetric in $\alpha,\beta$. 

For the other type of heavy neutrino propagator, only the first diagram exists (due to
$SU(2)_L$ charge considerations) and it has a different Yukawa index structure:
\begin{align}\label{chargeconserving}
&\Sigma_i(x',x) - \Sigma^c_i(x',x) \nonumber \\
& =    \sum_{\alpha, \, \beta, \, j}2i \mbox{Im}\left[ \lambda^\dagger_{\beta i}
  \lambda_{i\alpha} \lambda^\dagger_{\alpha j} \lambda_{j \beta}
\right]  G(x',x) \int d^d y \int d^d z \, \, G(y,z)
S_{[\alpha} (x',y)S_j(y,z) S_{\beta]}(z,x). \ 
\end{align}
As we see later, however, this does not contribute to the overall lepton number asymmetry. 

It is now clear that Eqs. (\ref{propagator}) , (\ref{vertex}) and (\ref{chargeconserving}) are non-vanishing in curved spacetime.
We see, therefore, that at two loops, leptons and antileptons propagate differently, due to a breaking of translation invariance 
by a general background, which is probed by curved-space Green functions, $S_j(x,y)$,  $G(x,y)$ and $S_\alpha(x,y)$.
In diagrammatic terms, this is how the time dependence necessary to evade the theorems in section \ref{flatpropagation}
arises in our mechanism for leptogenesis.
From a calculational point of view, the breaking of translation symmetry causes the two-loop self-energy
to become sensitive to the ordering of the sterile neutrinos, $N_\alpha$ and $N_\beta$, within it, which are distinguishable 
by virtue of their non-degenerate masses and provides an antisymmetric part to the Feynman integral. 

Given the arguments of section \ref{flatpropagation}, we must also find that, if we restore
translation invariance by going to Minkowski space, the differences (\ref{propagator})-(\ref{chargeconserving}) will vanish. 
Indeed, it is easy to check that by substituting flat space propagators, the difference in self-energies is zero. 
For instance, substituting flat space propagators into  (\ref{vertex}) gives (see  (\ref{flatprops}))
\begin{align}
&\Sigma(p, \eta_{\mu \nu}) - \Sigma^ c (p,\eta_{\mu \nu}) \nonumber \\
& =  \sum_{\alpha, \, \beta, \, j} 2i  \mbox{Im}\left[ \lambda^\dagger_{\beta i} \lambda_{i\alpha} \lambda^\dagger_{\beta j}\lambda_{j \alpha} \right]  
\nonumber \\
&\times \left( \int \frac{d^d k}{(2\pi)^n}  \int \frac{d^d \ell}{(2\pi)^n}  \frac{1}{ (k-p)^2 - m_H^2} \frac{1}{ (\ell-p)^2 - m_H^2} 
\frac{M_\alpha}{k^2 - M_\alpha^2}\frac{  \slashed{k} + \slashed{\ell} - \slashed{p}   }{ (k+\ell - p)^2} \frac{M_\beta}{\ell^2 - M_\beta^2}  
- (\alpha \leftrightarrow \beta) \right) \nonumber \\
&=0,
\end{align}
where, after a trivial relabeling of integration variables, the expression is easily seen to be symmetric under interchange of 
$M_\alpha$ and $M_\beta$, and hence zero. A similar result holds for the other diagrams. 

We now show how this difference in propagation manifests itself in the form of curvature-dependent, C and CP violating operators 
in the effective action generated by the diagrams in figure \ref{2loopagain}.

\section{Effective field theory - integrating out the sterile neutrinos}\label{EFTcurved}

One of the most direct ways to study the propagation of particles in gravitational backgrounds, and the effects of curvature 
on their dynamics, is to use effective field theory \cite{Drummond:1979pp, Ohkuwa:1980jx, McDonald:2014}. 
As discussed in the previous sections, the screening cloud surrounding an interacting particle gives it an effective size, 
causing it to experience tidal forces. When one integrates out the heavy sterile neutrinos, this phenomenon generates operators 
in the effective action that couple particle fields to various curvature tensors, suppressed by a see-saw scale cutoff. 
The most general such action \cite{McDonald:2014}, constructed from a complete basis of hermitian operators, 
is given to linear order in curvature and leading order in derivatives by
\begin{align}\label{eff1}
\mathcal{L}_{eff} = \sqrt{-g}   
\Bigg[   \bar{\ell}i \slashed{D} \ell &+ ia\bar{\ell}\left(2R_{\mu
    \nu}\gamma^\mu D^\nu   
+ \frac{1}{2}\partial_\mu R \gamma^\mu \right) \ell + b \partial_\mu R  \bar{\ell} \gamma^\mu \ell\nonumber\\
&+ ic \bar{\ell}\left( 2R \slashed{D}  +  \partial_\mu R \gamma^\mu
\right) \ell + id \bar{\ell}\left( 2 D^2\slashed{D}+ \frac{1}{4}\partial_\mu R
  \gamma^\mu \right)\ell \Bigg] ,
\end{align} 
where $a,b,c,d$ are real, effective couplings, with mass dimension minus
two, which will depend on $\lambda_{i\alpha}$ and the masses $M_\alpha$ and $M_\beta$ in the loops. 
There is one term in this effective action which is of great importance for leptogenesis, and is the only C and CP violating 
operator in (\ref{eff1}), {\it viz.}
\begin{equation}
\mathcal{L}_{CPV} = b \, \partial_\mu R  \, \bar{\ell} \gamma^\mu
\ell \label{dmuR} \ .
\end{equation}
A careful discussion of the action of C, P and T on this, and remaining operators appearing in $\mathcal{L}_{eff}$ 
was given in \cite{McDonald:2014}. 

The presence of a C and CP violating operator ensures that the second Sakharov condition holds, 
namely that the complete theory contains C and CP violation, which is provided here by the 
complex couplings $\lambda_{i \alpha}$. 
In our model, this means that the operator will be generated at the two-loop level by precisely the same diagrams 
which lead to asymmetric propagation of matter and antimatter, and will depend on the complex phases contained 
in the Yukawa couplings $\lambda_{i \alpha}$. In the next section, we show how integrating out the heavy neutrinos
from these diagrams gives an expression for the effective coupling $b$, whose calculation we shall now describe in detail. 

At this point, we should comment further on the range of validity of the effective Lagrangian (\ref{eff1})
in the context of leptogenesis. It is written to first order in the curvatures, so is valid for small values of the 
parameter ${\frak R}/M^2$, where ${\frak R}$ denotes a typical curvature component while $M$ is the heavy scale,
provided here by the sterile neutrino masses. It is also a low-energy Lagrangian, keeping only terms of 
leading order in derivatives. As shown in the series of papers \cite{Hollowood:2008kq, Hollowood:2009qz, Hollowood:2011yh} 
which discuss the realisation
of causality and the energy dependence of theories of this type, the relevant parameter here is $E \sqrt{{\frak R}}/M^2$, 
where $E$ denotes the lepton energy scale. Both these parameters are required to be small for the validity of the effective 
Lagrangian. We return to this point in section \ref{leptogenesis}, where we apply
the effective Lagrangian and chemical potential to the hot, early Universe where both $E$ and ${\frak R}$ are
related to the temperature.

\subsection{Matching}

\begin{figure}[h!]
\centering
\includegraphics[scale=0.6]{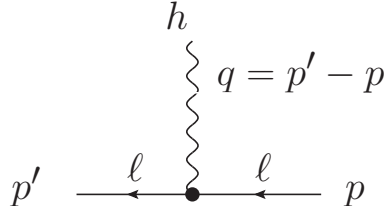}
\caption{The effective $h$ vertex generated by $\mathcal{L}_{CPV}$ where $q=p'-p$ 
is the momentum transfer between the ingoing and outgoing lepton.}
\label{effective}
\end{figure}

The calculation of the effective couplings in (\ref{eff1}) is performed in the usual way, by matching 
the fundamental UV-complete theory to the effective action at low energy. In particular, the calculation can be reduced 
to the problem of evaluating a certain class of two-loop Feynman diagrams, which we now describe.  
The first step is to notice that the effective couplings are independent of the choice of background, 
so that a judicious choice of metric greatly simplifies the computation. We shall pick a conformally flat metric 
\begin{equation}\label{metric}
g_{\mu \nu} = \Omega^2 \eta_{\mu \nu} = (1 + h)\eta_{\mu \nu} \ ,
\end{equation}
which is sufficient to distinguish the various components of the effective Lagrangian (\ref{eff1}). 
In particular, since the Ricci tensor is given by
\begin{equation}
R_{\mu \nu} = - \partial_\mu \partial_\nu h - \frac{1}{2} \eta_{\mu \nu} \partial^2 h + \mathcal{O}(h^2),
\end{equation}
we see that the effective couplings can be determined by working to linear order in $h$. 
For instance, since $R = - 3 \partial^2 \Omega^2$, the contribution to the effective vertex from the 
operator $\mathcal{L}_{CPV} = b \partial_\mu R \bar{\ell} \gamma^\mu \ell$,  is given by
\begin{equation}\label{LCPV}
\mathcal{L}_{CPV} =-3 b \left(\partial_\mu \partial^2 h \right) \bar{\ell} \gamma^\mu \ell + O(h^2).
\end{equation}
We can then use the Minkowski background to define a momentum space, over which $h$ is treated 
as a classical background field. This gives a contribution from this operator of the form
\begin{equation}\label{eff}
A(q) = 3 i b (q^2 \slashed{q}) \, h(q),
\end{equation}
which corresponds to the diagram in figure \ref{effective}.

\noindent Similarly we can expand the Lagrangian of the UV theory 
\begin{align}\label{seesaw}
\mathcal{L} = \mathcal{L}_{EW} + \sqrt{-g}\left[ \overline{N} i\slashed{D} N+ \lambda_{i \alpha} \bar{\ell}_i\phi  N_\alpha   
+ \frac{1}{2}\overline{\left(N^c\right)} 
\, M \,  N  + \mbox{h.c.}\right]  \ ,
\end{align} 
to linear order in $h$. The computation is also simplified if we work with conformally rescaled fields, 
\begin{equation}
N\rightarrow  \Omega^{-(n-1)/2}N, \quad   \ell  \rightarrow  \Omega^{-(n-1)/2}\ell,   \quad \phi 
\rightarrow \Omega^{-(n-2)/2}\phi .
\end{equation}
After conformal rescaling of the Lagrangian (\ref{seesaw}) and inserting the metric (\ref{metric}), 
gravity enters only via the terms which violate conformal invariance, so that the $\Omega$-dependent terms
in the Lagrangian can be written as
\begin{align}
\mathcal{L}_{\Omega} &=  \frac{1}{2}\Omega \overline{N^c} M N +
\Omega^2  \left(  m_H^2 - 6 \left(\zeta - \frac{1}{6}\right)\Omega^{-3} \partial^2 \Omega \right)\phi^\dagger \phi \, + \,
\Omega^{-(n-4)/2}\lambda_{i \alpha} \bar{\ell}_i \phi N \  \nonumber \\
&=   \mathcal{O}h + O(h^2),
\end{align}
where the Higgs Lagrangian includes the Ricci coupling $\zeta R \phi^\dagger \phi$.
Expanding to linear order in $h$, we have
\begin{align}\label{hinsertion}
\mathcal{O}h= \frac{1}{4} h M \overline{N} N^c + \mbox{h.c.} +  
\left(  m_H^2 h - 3\left(\zeta - \frac{1}{6}\right)\partial^2 h  \right)\phi^\dagger \phi \, -\frac{(n-4)}{4} h \lambda_{i \alpha} \bar{\ell}_i \phi N  .
\end{align}
In this way, gravity manifests itself in the form of a classical background field $h$, so that the two-loop diagrams 
can be expanded to linear order in $h$ by appropriate insertion of $h$ according to the operators in (\ref{hinsertion}). 
This reduces the problem to the evaluation of flat space 3-point Feynman diagrams, with two external fermion legs 
and a classical field $h$. 

\begin{figure}
\centering
\includegraphics[scale=0.54]{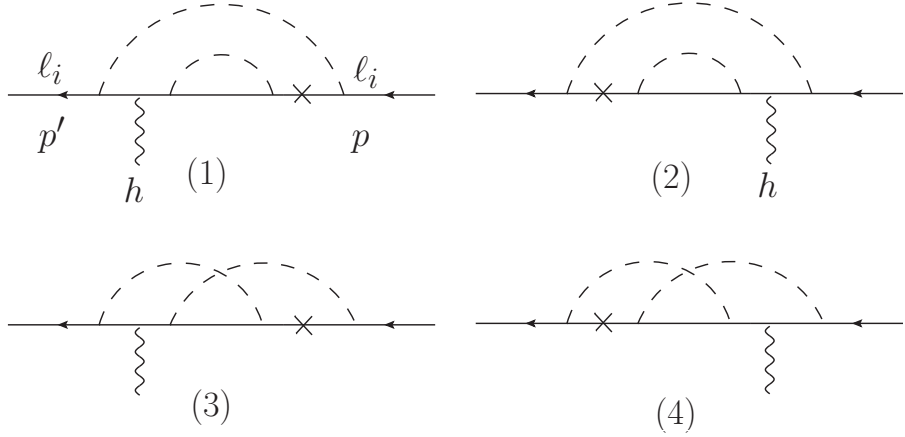}
\caption{Contributions to $\bra{\,\ell(p') \,}  \mathcal{O}h \ket{\, \ell(p)\,}$ from the heavy neutrino mass term 
$\frac{1}{2}h\bar{N} M N^c$. The crosses denote the $S^\times_\beta$ sterile neutrino propagator, while on the lines
with $h$ insertions
there are contributions $S_\alpha S_\alpha$ and $S_\alpha^\times S_\alpha^\times$  corresponding to each propagator type.}
\label{2loops}
\end{figure}

For the case of $h$ couplings to the heavy neutrinos via their mass term, the corresponding 
diagrams are shown in figure \ref{2loops}.
Suppose that the $h$ insertion is made into the $N_\alpha$ propagator.  Recalling that only the diagrams with both
propagators of the charge-violating type contribute to leptogenesis, the $N_\beta$ propagator must be of the type 
$S^\times_\beta$ in (\ref{flatprops}). 
Then, given the two terms in (\ref{hinsertion}) for the coupling of $h$ to the sterile neutrinos, 
{\it viz.} $\frac{1}{4} h M \left(\overline{N} N^c + \overline{N^c}N\right)$,
we see that with this condition the $N_\alpha$ line receives contributions from {\it both} $S_\alpha S_\alpha^c$
and $S^\times_\alpha S^\times_\alpha$. Finally, we use $S_\alpha^c = S_\alpha$ in flat space.
This establishes the form of the diagrams to be calculated in the following section.

The effective couplings can be computed by matching the transition matrix elements 
$\bra{\,\ell(p') \,}  \mathcal{O} h \ket{\, \ell(p)\,}$ for small external momenta to the effective amplitudes 
such as (\ref{eff}) (see in particular \cite{Ohkuwa:1980jx, McDonald:2014}, as well as 
\cite{Drummond:1979pp, Berends:1975ah, Donoghue:2001qc,
  BjerrumBohr:2002ks}, for more details). 
The general form of this object is
\begin{equation}
\bra{\,\ell_i(p') \,}  \mathcal{O}h \ket{\, \ell_i(p)\,} = \slashed{p} \left[ \alpha_1 p^2 
+ \alpha_2 (p\cdot q) + \alpha_3 q^2 \right] + 
\slashed{q} \left[ \beta_1 p^2 + \beta_2 (p\cdot q) +\beta_3 q^2\right]
\end{equation}
where $q=p'-p$ is the momentum transfer, and $\alpha_i$ and $\beta_i$ are in general complex coefficients, 
which depend on the Yukawa couplings and the masses in the loop. From (\ref{eff}), we see that the coefficient 
$b$ can be read off as $b = 1/3 \, \mbox{Im}(\beta_3)$, which in turn only depends on the value of the 
momentum transfer $q$. Hence, for the purposes of calculating the operator (\ref{dmuR}), we can set $p=0$ 
in the remainder of our calculations, so that the amplitudes are functions of a single momentum $q$, 
with $p'=q$. The transition amplitude is thus given by
\begin{align}\label{pOp}
\bra{\,\ell_i(q) \,}  \mathcal{O}h \ket{\, \ell_i(0)\,} &= \sum_{\alpha,  \,  \beta, \,
  j} \lambda^\dagger_{\beta i} \lambda_{i\alpha}\lambda^\dagger_{\beta j} \lambda_{j \alpha}  \, f(q^\mu,M_\alpha,M_\beta)   
\nonumber \\
& = i q^2 \slashed{q} \, h(q)\sum_{\alpha,  \,  \beta, \,
  j} \frac{\mbox{Im}\left[ \lambda^\dagger_{\beta i} \lambda_{i\alpha}
    \lambda^\dagger_{\beta j} \lambda_{j \alpha} \right] }{M_\alpha
  M_\beta}I_{[\alpha \beta]} + \cdots,
\end{align}
where $+ \cdots$ represents terms which do not contribute to (\ref{eff}) and $I_{[\alpha \beta]}$ is antisymmetrised
on $\alpha,\beta$.
 The factor $I_{\alpha \beta} = I (M_\alpha ,M_\beta)$ depends on the masses of the sterile neutrinos. It can be determined 
by performing a momentum expansion of $f(q,M_\alpha,M_\beta)$, in the limit $-q^2 \ll M_\alpha, M_\beta$, 
where $f$  is directly determined from the evaluation of the three-point Feynman diagrams in figure \ref{2loops}. 
In this sense, the see-saw scale becomes the UV cut-off for our effective theory. Matching the effective 
amplitude (\ref{eff}) to (\ref{pOp}) we find that
\begin{equation}\label{CPVoperator}
\mathcal{L}_{CPV}=\partial_\mu R \, \bar{\ell}_i \gamma^\mu \ell_i \,
\sum_{\alpha,  \,  \beta, \,   j} \frac{\mbox{Im}\left[
    \lambda^\dagger_{\beta i} \lambda_{i\alpha} \lambda^\dagger_{\beta
      j} \lambda_{j \alpha} \right]}{3 M_\alpha M_\beta} I_{[\alpha \beta] }.
\end{equation}
In this way, we have shown how this operator arises simply from curved space QFT considerations, 
without the need to postulate new gravitational interactions arising from some unspecified, more fundamental theory.  

\begin{figure}[h!]
\centering
\includegraphics[width=7.2cm,height=5cm]{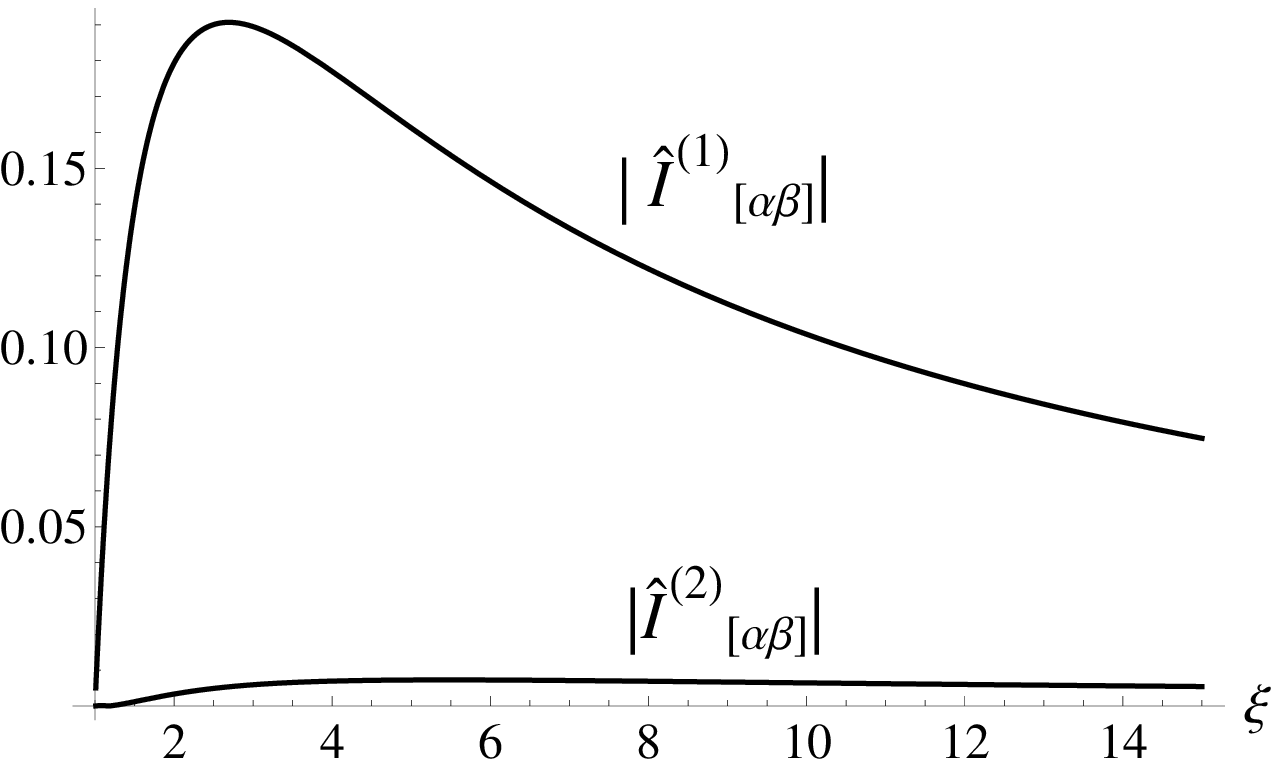} \hskip0.5cm
\includegraphics[width=7.2cm,height=5cm]{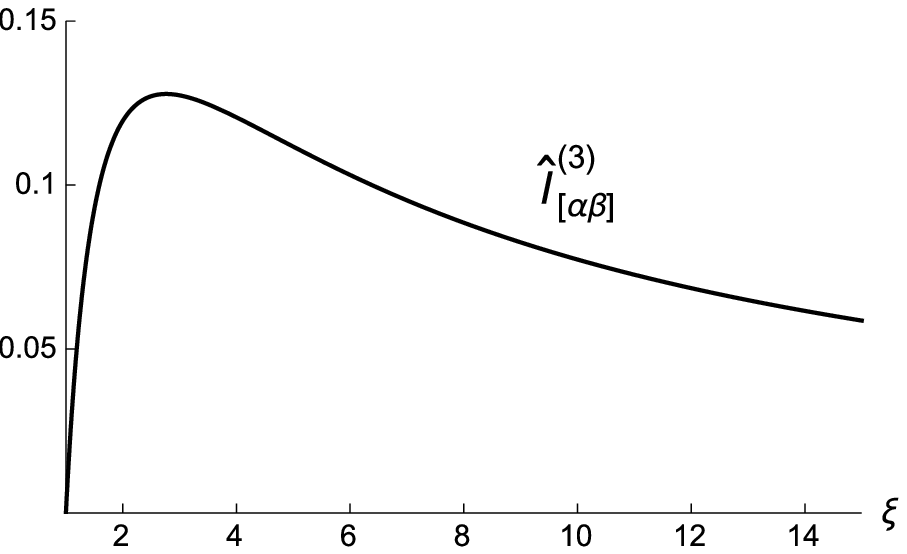}
\caption{The dependence of the two-loop
  Feynman diagrams (1) to (3) on the sterile neutrino masses. 
The curves show
$\hat{I}_{[\alpha\beta]}^{(n)} = (4\pi)^4 |I_{[\alpha\beta]}^{(n)}|/\xi$ plotted as a function of the heavy sterile neutrino
mass ratio $\xi = M_\beta/M_\alpha$ with $M_\alpha$ fixed.} 
\label{Iab}
\end{figure}

We describe the relevant Feynman diagram calculations for the coefficient $I_{\alpha \beta}$ at some length in the next section
and show explicitly that it develops the antisymmetric part necessary to generate the operator (\ref{CPVoperator}).
The full expressions for  $I_{[\alpha \beta]}$ may be expressed in terms of the mass ratio 
$\xi = M_\beta/M_\alpha$ of the sterile neutrinos
and the results for diagrams (1) to (3) are plotted in figure \ref{Iab}.
For a large hierarchy, $\xi \gg 1$ with $M_\alpha$ fixed, we find the leading behaviour
\begin{equation}\label{I12limit}
\frac{I_{[\alpha \beta]}^{(1)}}{M_\alpha M_\beta} \simeq  - \frac{5}{16}\frac{1}{(4\pi)^4}\frac{1}{M_\alpha^2 } \frac{1}{\xi} \ , 
\qquad \qquad 
\frac{I_{[\alpha \beta]}^{(2)}}{M_\alpha M_\beta} \simeq  - \frac{1}{48}\frac{1}{(4\pi)^4}\frac{1}{M_\alpha^2 } \frac{\ln \xi^2}{\xi} \ ,
\end{equation}
for the ``propagator correction'' diagrams (1) and (2), and
\begin{equation}
\frac{I_{[\alpha \beta]}^{(3)}}{M_\alpha M_\beta}   \simeq   \frac{1}{12}\frac{1}{(4\pi)^4}\frac{1}{M_\alpha^2 } 
\frac{\log \xi^2}{\xi} \ , 
\label{hierarchy3}
\end{equation}
for the ``vertex correction'' diagram (3).
The final diagram (4) is significantly more difficult to calculate than the others and the issues involved
are explained below.

These 2-loop calculations, even in the low-momentum limit, are not simple, and considerable care 
must be taken in particular to deal with the various massless thresholds which arise. Note that it is only
the terms of $O(q^2 \slashed{q})$ which contribute to the local effective Lagrangian (\ref{LCPV}). 
We also encounter non-analytic terms involving $\ln(-q^2)$, which are to be interpreted as
non-local contributions to the effective action. While such terms are of considerable importance in their
own right and encode important information about the long-range interactions in the theory
\cite{Donoghue:2001qc, BjerrumBohr:2002ks}, they do not affect the leptogenesis mechanism of interest to us here.
We now discuss these calculations in detail, but the reader interested primarily in the 
implications for cosmology may at this point jump ahead directly to section \ref{leptogenesis},
where the consequences for leptogenesis are discussed.

\section{Feynman diagram calculations}\label{Results}

In this section we describe the calculation of the two-loop self-energy diagrams diagrams shown in figure \ref{2loops}. 
Diagrams (1)-(3) can be evaluated by first evaluating the one-loop propagator or vertex sub-diagrams,
which in these cases are relatively simple, then inserting into the full self-energy diagrams.  
However, the sub-diagram for (4) is of a non-trivial triangle type \cite{'tHooft:1978xw} 
and here the whole diagram must be dealt with in a different way.

\subsection{Sterile neutrino couplings}

First, we describe the diagrams where the gravitational field $h$ couples to the heavy sterile neutrino propagators.

\section*{Diagram (1)}

\begin{figure}[h]
\centering
\includegraphics[scale=0.5]{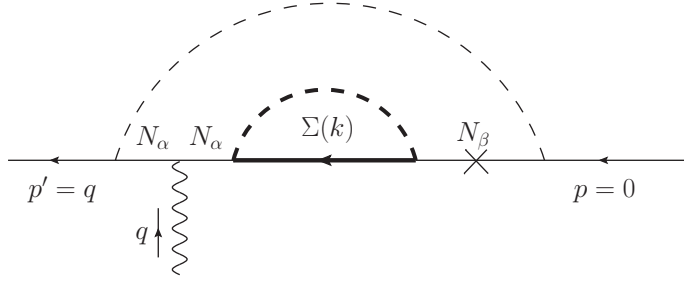}
\caption{The propagator correction diagram with no branch cuts. The self-energy sub-diagram $\Sigma(k)$ 
is shown in bold and we sum over the two kinds of sterile neutrino propagators which can be inserted at the $h$ vertex. }
\label{N1}
\end{figure}
\noindent This diagram gives
\begin{align}
& f^{(1)} (q^\mu, M_\alpha,M_\beta) = \frac{M_\alpha}{2} \int \frac{d^d k }{(2\pi)^d} \, 
G(k) \left[ S_\alpha (k+q) S_\alpha (k) + S_\alpha^\times(k+q)S^\times_\alpha(k)\right] \Sigma(k) S^\times_\beta (k) , \nonumber \\ 
\end{align}
where $\Sigma(k)$ is the massless sub-diagram shown in bold in figure \ref{N1}:
\begin{equation}
\Sigma(k) = \int \frac{d^d \ell}{(2\pi)^d} G(k - \ell) S(\ell). 
\end{equation}
It is easily evaluated to give 
\begin{equation}
\Sigma(k) = \frac{i\slashed{k}}{(4\pi)^2} \left[ \frac{1}{2}\ln \left(- \frac{k^2}{\mu^2}\right) - 1\right],
\end{equation}
where $\mu$ is an RG scale, and we used $\overline{MS}$ when removing the pole. One can then re-insert this
 into the main diagram to find $f^{(1)}(q,M_\alpha,M_\beta)$, which can be written as a momentum expansion in $q$. 
The details of this expansion and the techniques necessary to evaluate the full diagram are described in the appendix. 
One can then read off the contribution to $I_{\alpha \beta}$ and its dependence on the masses $M_\alpha$ and $M_\beta$. 
For this diagram, we find a contribution to $I_{\alpha \beta}$ given by
\begin{align}\label{I1}
I^{(1)}_{\alpha \beta} = \frac{1}{{24 (4\pi)^4 (x-1)^5}}\Big[& 24 x \ln (x) \left(\left(x^2+x-1\right) \ln\left(\frac{\mu^2}{M_\alpha^2}\right)+x\right)\nonumber \\
& +2 (x-1) (x (x ((x-5) x-19)+7)+4) \ln \left(\frac{\mu^2}{M_\alpha^2}\right) \nonumber \\
& -12 \left(x^2+x-1\right) x
  \ln^2 x  -3 x^5+28 x^4-54 x^3+41 x-12 \Big]
\end{align} 
where
\begin{equation}
x = \xi^2 = \frac{M_\beta^2}{M_\alpha^2} .
\end{equation}
The large $x$ behaviour for the antisymmetric part, choosing the RG scale $\mu=M_\alpha$, is therefore given by
\begin{equation}
I_{[\alpha \beta]}^{(1)} \simeq  - \frac{5}{16} \frac{1}{(4\pi)^4}+ O\left(\frac{\ln^2 x}{x}\right), \qquad  x \gg 1 \ ,
\end{equation}
as quoted in (\ref{I12limit}).

\section*{Diagram (2)}

\begin{figure}[htbp]
\centering
\includegraphics[scale=0.5]{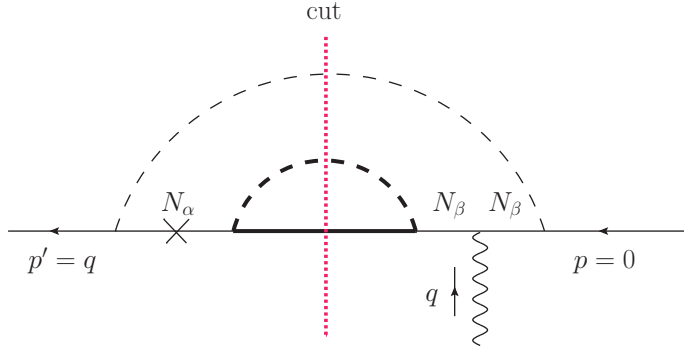}
\caption{The propagator correction diagram with a zero-mass threshold corresponding to the cut (shown in red) through the 
massless Higgs and lepton lines, across which the momentum $q$ flows.}
\label{N2}
\end{figure}

This diagram contains a zero-mass threshold \cite{Smirnov et al} as shown by the cut in figure \ref{N2}. 
As a result, the diagram has a discontinuity at $q^2=0$ and a branch cut represented by the appearance 
of $\ln\left(\frac{-q^2}{M_\alpha^2}\right)$ in the expansion for  $f^{(2)}(q,M_\alpha,M_\beta)$. We find
\begin{equation}
f^{(2)}(q,M_\alpha,M_\beta)=\frac{\slashed{q}}{M_\alpha M_\beta}\left[ I^{(2)}_{\alpha \beta} -\frac{1}{24} 
\ln \left( -\frac{q^2}{M_\alpha^2}\right)  \right]q^2 + \cdots,
\end{equation}
with $I_{\alpha \beta}^{(2)}$ given by
\begin{align}
I^{(2)}_{\alpha \beta}=\frac{1}{144 (4\pi)^4 (x-1)^5}\Big[& 6 \ln \left(\frac{\mu^2}{M_\alpha^2} \right) 
\left(2 x^5-11 x^4+28 x^3-20 x^2-12 x^2 \ln (x)+2 x-1\right) \nonumber \\
& +16 x^5-119 x^4+88 x^3-34 x^2+36 x^2 \ln ^2(x) \nonumber \\
& -6 \left(x^4-6 x^3-6 x^2-10 x-3\right) x \ln (x)+56 x-7\Big]
\end{align}
In the large hierarchy limit, the antisymmetric part in this case is
\begin{equation}
I_{[\alpha \beta]}^{(2)} \simeq  - \frac{1}{48} \frac{\ln x }{(4\pi)^4} + O(1),  \qquad  x \gg 1\ .
\end{equation}

\section*{Diagram (3)}

\begin{figure}[h]
\centering
\includegraphics[scale=0.5]{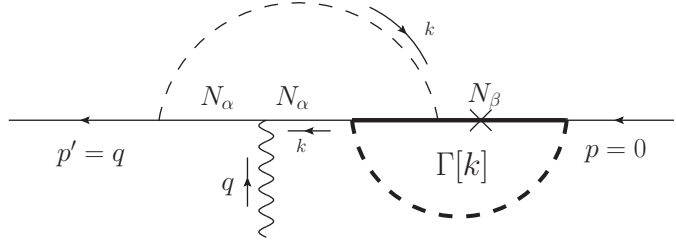}
\caption{The vertex-correction diagram with no branch cuts. The vertex correction sub-diagram $\Gamma(k)$, 
is shown in bold, with a single momentum $k$ flowing in and back out. }
\label{N3}
\end{figure}

As before, the relevant subdiagram is shown in bold (figure \ref{N3}).
Although this sub-diagram is of the triangle type \cite{'tHooft:1978xw} which, in general, gives a lengthy answer, 
since there is only one momentum flowing into it, and since it contains only one mass, the expression is relatively simple:
\begin{equation}
\Gamma(k) = 
 -\frac{i \slashed{k}}{(4\pi)^2 M_\beta}
 \left\{ \left[ \frac{M_\beta^4}{k^4}- 2\frac{M_\beta^2}{k^2} \right]\ln\left( \frac{M_\beta^2 - k^2}{M_\beta^2}\right)
+  \frac{M_\beta^2}{k^2} 
+  \ln\left( \frac{k^2 - M_\beta^2}{k^2}\right) 
\right\}.
\end{equation}
Substituting this into the main diagram, one can evaluate the remaining $k$ integral following the procedure
described in the appendix.  There are no branch cuts in this diagram, and hence no $\ln(-q^2)$ terms. 
The final contribution is\footnote{We are grateful to T. Shindou and S. Shirai for pointing out an error in our earlier calculation
of diagram (3), published in JHEP 1604 (2016) 030, which resulted in a different sterile neutrino mass
dependence in (\ref{I3hierarchy}).}
\begin{align}
I^{(3)}_{\alpha \beta}  = \frac{x}{6 (4\pi)^4 (x-1)^4} \Bigg[ 
&\left(2\pi^2 x^4 - 2(6 + \pi^2)x^3 + (15-6\pi^2)x^2 + 10(2+\pi^2) x - 4\pi^2 -35\right) (x-1)    \nonumber \\
&+\left(-12x^4 + 19 x^3 + 28 x^2 - 73 x + 50\right) \log x     \nonumber \\
&- 12 (x+2) (x-1)^4 \, {\rm Li}_2\left(\frac{x-1}{x}\right) \Bigg].
\end{align}
We are now in a position to derive the result (\ref{hierarchy3}). After antisymmetrisation, we find that the 
asymptotic behaviour of this diagram is given by
\begin{equation}
I^{(3)}_{[\alpha \beta]} \simeq \frac{1}{12} \frac{\ln x}{(4\pi)^4}  + O(1) \ ,\qquad  x \gg 1.
\label{I3hierarchy}
\end{equation}

\section*{Diagram (4)}

\begin{figure}[h]
\centering
\includegraphics[scale=0.5]{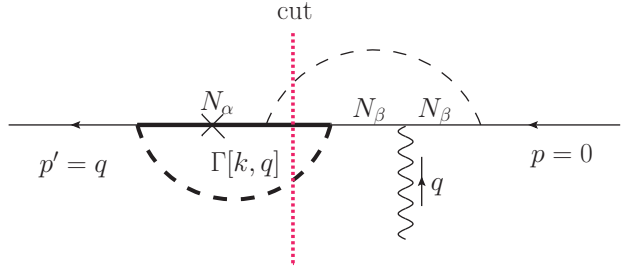}
\caption{The vertex-correction diagram with a branch cut through Higgs and lepton lines.}
\label{N4}
\end{figure}

The amplitude for this process is given by
\begin{align}
f^{(4)}(q,M_\alpha,M_\beta)   &= \frac{ M_\beta}{2}  \int \frac{d^d k}{(2\pi)^d}\int  \frac{d^d \ell}{(2\pi)^d}   
\frac{M_\alpha (\slashed{k}+\slashed{\ell})[M_\beta^2 + \slashed{k} (\slashed{k} - \slashed{q}) ]}{k^2[ k^2 - M_\beta^2 ] 
\ell^2   \left[ (k+\ell)^2 \right]  [ (k-q)^2 - M_\beta^2 ]\left[ (\ell+q)^2 - M_\alpha^2 \right]}, \nonumber \\
\end{align}
and corresponds to the diagram in figure \ref{N4}.

\begin{figure}[h]
\centering
\includegraphics[scale=0.5]{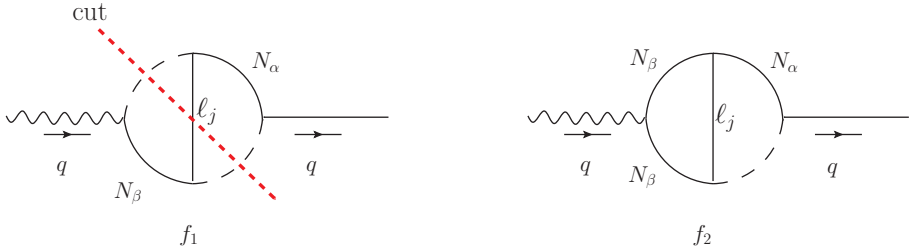}
\caption{Self-energy diagrams resulting from partial fractioning.}
\label{selfE}
\end{figure}

For this amplitude, the sub-diagram (shown in bold) is also of the triangle type. 
However in this instance it depends on two momenta, rather than just one. This vastly complicates the form 
of the sub-diagram, which is worthy of a separate analysis \cite{'tHooft:1978xw} in its own right. 
Furthermore, the overall diagram contains a branch cut, shown in red. Not only that, but the complicated sub-diagram 
has to be substituted into the remaining momentum integral. Since we only require a momentum expansion, 
we shall proceed in a different way from the previous diagrams. First, we notice that we can use partial fractions to write
\begin{equation}
\frac{1}{k^2}\cdot\frac{1}{k^2 - M_\beta^2} =- \frac{1}{M_\beta^2} \left(  \frac{1}{k^2}- \frac{1}{k^2 - M_\beta^2 } \right)
\end{equation}
which allows the diagram to be written as the difference of two self-energies of the types shown in figure \ref{selfE}. 
This gives the decomposition,
\begin{equation}
f^{(4)}(q,M_\alpha,M_\beta) = -\frac{1}{M_\beta^2}\left[ f_1(q,M_\alpha,M_\beta)  -f_2(q,M_\alpha,M_\beta) \right].
\label{4decomp}
\end{equation}

This class of diagrams has been extensively studied in the literature \cite{Smirnov et al,Tausk1994,Davydychev}. 
The first step is to simplify the gamma matrix expression in the numerator and write as far as possible 
as a sum of terms appearing in the propagator denominators. This allows the full diagram to be written
as a sum of scalar diagrams with a smaller number of propagators, as shown below. Some of these are UV divergent,
but the poles are all spurious, in the sense that their sum must be finite, 
as the overall diagram (4) contains no divergences.  For instance, in the case of $f_1$, we find the diagram decomposes as
\begin{align}
\label{scalar}
f_1=    \frac{\slashed{q}}{2}  \Bigg[&  \frac{1}{2 q^2} \frac{M_\alpha}{M_\beta}\Big( J( 0,1,1,1,0) - J(0,1,1,0,1) - J(0,0,1,1,1) + J(-1,1,1,1,1)\Big)  
\nonumber \\
& + \frac{M_\alpha}{M_\beta}\left(  \frac{ M_\beta^2+ M_\alpha^2}{2 q^2}  -\frac{1}{2} \right) J(0,1,1,1,1) + \frac{M_\alpha}{2 M_\beta} J(1,0,1,1,1)   
- \frac{M_\alpha}{2 M_\beta} J(1,1,0,1,1)
\vspace{0.2cm}  \nonumber \\
&- \frac{M_\alpha M_\beta}{q^2} \left[ J(1,1,1,0,1) +J(1,0,1,1,1) \right] + \frac{M_\alpha M_\beta}{q^2}[ J( 1,1,1,1,0)+J( 0,1,1,1,1)]  \nonumber \\
&+ \left(\frac{(M_\alpha^2 + M_\beta^2 )}{ q^2}  + \frac{1}{2} \right) M_\alpha M_\beta  J(1,1,1,1,1)  \Bigg] \ ,
\end{align}
where
\begin{align}
&J(\nu_1,\nu_2,\nu_3,\nu_4,\nu_5) \nonumber \\
&= \int \frac{d^d k_1}{(2\pi)^d}\int \frac{d^d k_2}{(2\pi)^d}
\frac{1 }{ [k_1^2- M_\beta^2]^{\nu_1}\left[ k_2^2\right]^{\nu_2}  \left[ (k_1+k_2)^2 \right]^{\nu_3}  
\left[ (k_1-q)^2 \right]^{\nu_4}\left[ (k_2+q)^2 - M_\alpha^2 \right]^{\nu_5}}.
\end{align}
\begin{figure}[h!]
\centering
\includegraphics[scale=0.45]{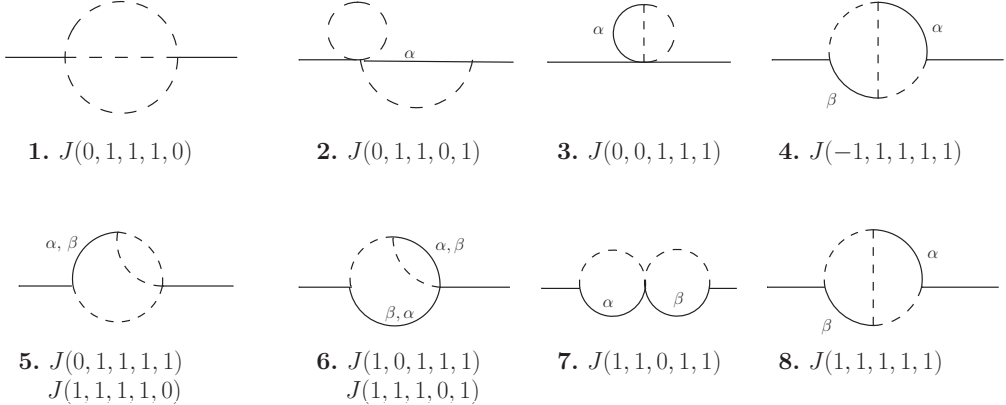}
\caption{The scalar topologies for $f_1$. The dashed lines now generically denote massless propagators, 
and so can correspond to both Higgs and lepton lines. }
\label{K}
\end{figure}
The corresponding diagrams are given in figure \ref{K}.
Not all these diagrams give a $q^2 \slashed{q}$ contribution. Firstly, we note that diagram $2$ is zero, 
since it contains a massless bubble, which vanishes in dimensional regularisation. Futhermore, diagram 3 
has no $q$ dependence, and so does not contribute to $q^2 \slashed{q}$. We also see that last 2 lines of (\ref{scalar}) give 
contributions which are symmetric under interchange of $\alpha$ and $\beta$, and so do not contribute to $I_{[\alpha \beta]}$. 
Hence there are only 5 diagrams which contribute: 1, 4, 5, 6 and 7. Diagrams 1 and 5 were computed by Scharf and Tausk 
\cite{Tausk1994} and yield\footnote{In the notation of \cite{Tausk1994} corresponds to $T_{1234}(q^2;M_\alpha^2,0,0,0)$ 
of equation (96).  }
\begin{align}
J(0,1,1,1,0)& = - \frac{1}{(4\pi)^4}\left[ \frac{1}{4\delta} - \frac{13}{8} + \frac{1}{2}
 \left\{\gamma + \ln\left(\frac{-q^2}{4 \pi \mu^2}\right) \right\}  \right].\\
\nonumber \\
J(0,1,1,1,1) &=  -\frac{1}{(4\pi)^4} \Bigg[ \frac{1}{2 \delta^2} + \frac{1}{2 \delta} 
\left \{ 5 - 2 L_\alpha - 2 \left( \frac{1+Q}{Q} \right)\ln (1+Q)  \right\}\nonumber \\
& + \frac{19}{2} + \frac{3}{2} \zeta(2) + L_\alpha^2 - L_\alpha \left(5 - 2 \left(\frac{1+Q}{Q} \right) \ln(1+Q)\right)\nonumber \\
& -\ln\left( Q\right) + \left(\frac{1+Q}{Q} \right) \Big\lbrace 2 \mbox{Li}_2 (-Q) +\ln^2 (1+Q) \nonumber \\
&+ \ln (Q) \ln(1+Q) - 4 \ln (1+Q) \Big \rbrace \Bigg] .
\end{align}
where
\begin{equation}
Q = - \frac{q^2}{M_\alpha^2}, \qquad L_\alpha = \gamma + \ln \left(\frac{M_\alpha^2}{4 \pi \mu^2} \right), \qquad d = 4 - 2 \delta.
\end{equation}
The corresponding result for $J(1,1,1,1,0)$ is given by replacing $M_\alpha$ with $M_\beta$. We also find that
\begin{align}
J(1,0,1,1,1) & =- \frac{1}{(4\pi)^4} \Bigg[ -\frac{Q}{2 x } \frac{1}{\delta} +\frac{Q }{6 (x-1) x^2} 
\bigg \lbrace 3 x^2 L_{\alpha }-9 x^2-3 x L_{\alpha }-6 x \text{Li}_2(1-b) \nonumber \\
& +6 \text{Li}_2(1-x)+\pi ^2 x+9 x-3 x \ln x-\pi ^2 \bigg \rbrace \Bigg] + \mathcal{O}\left( \frac{q^4}{M_\alpha^4}\right)
\end{align}
and
\begin{equation}
J(1,1,0,1,1) =K(q^2,M_\alpha)K(q^2,M_\beta)
\end{equation}
where
\begin{equation}
 K(q^2, M)= \frac{1}{\delta } - L_M + 2+  \ln \left(\frac{M^2}{ M^2 - \ell^2} \right) -\frac{M^2}{\ell^2} 
\ln \left( \frac{M^2}{M^2 - \ell^2}\right)
\end{equation}
we also have
\begin{equation}
J(-1,1,1,1,1)=I + (q^2 - M_\beta^2)J(0,1,1,1,1)
\end{equation}
where
\begin{equation}
I =\frac{M_{\alpha }^2}{(4\pi)^4} \left\{\frac{1}{9} Q^2  (4-3 \ln (Q))-\frac{1}{96} Q 
\left(24 L_m^2-84 L_m+10 \pi ^2+105\right)\right\}  + \mathcal{O}\left( \frac{q^6}{M_\alpha^6}\right).
\end{equation}

\begin{figure}
	\centering
\includegraphics[scale=0.8]{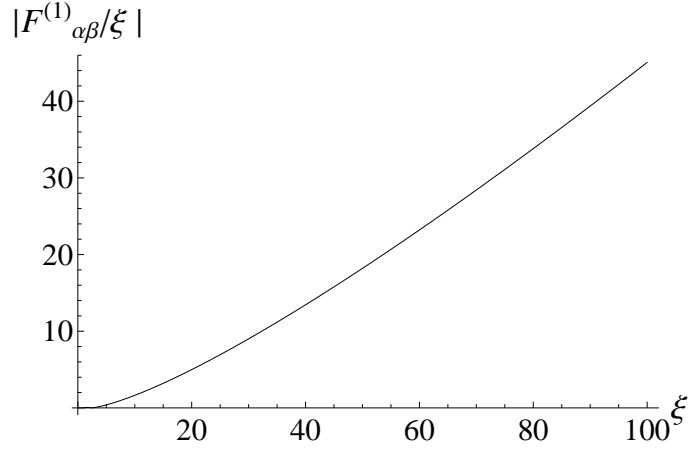}
\caption{The sterile neutrino mass dependence of the contribution to diagram (4)
from the $f_1$ term, where $\xi = M_\beta/M_\alpha$ with $M_\alpha$ fixed. As in figure \ref{Iab}, 
we have taken out an overall factor $1/(4\pi)^4$.} 
\label{F1}
\end{figure}

Putting all this together, we find a total contribution to $I^{(4)}$ from $f_1$ of the the form
\begin{equation}
f_1 (q,M_\alpha, M_\beta ) = \frac{\slashed{q} q^2}{M_\alpha M_\beta} F^{(1)}_{[\alpha \beta]} + \cdots,
\end{equation}
where 
\begin{align}
F^{(1)}_{[\alpha \beta]} =\frac{1}{48 (4\pi)^4 x^2} \Bigg[&6 x^4 \text{Li}_2\left(\frac{x-1}{x}\right)-
6 \text{Li}_2(1-x)-\pi ^2 x^4+12 x^3 \nonumber \\
& +3 x^3 \ln(x)+3 x^2 \ln (x)-12 x+3 x \ln (x)+\pi ^2 \Bigg].
\end{align}
It has the asymptotic behaviour (characterised by $\xi = M_\beta/M_\alpha$)
\begin{equation}
\frac{F^{(1)}_{[\alpha \beta]}}{M_\alpha M_\beta} \simeq - \frac{1}{16}\frac{1}{(4\pi)^4} \frac{1}{M_\alpha^2}\xi  \ln{\xi^2} , 
\qquad \xi \gg 1,
\label{F1hierarchy}
\end{equation}
as shown in the plot in figure \ref{F1}.

\begin{figure}[h!]
\centering
\includegraphics[scale=0.4]{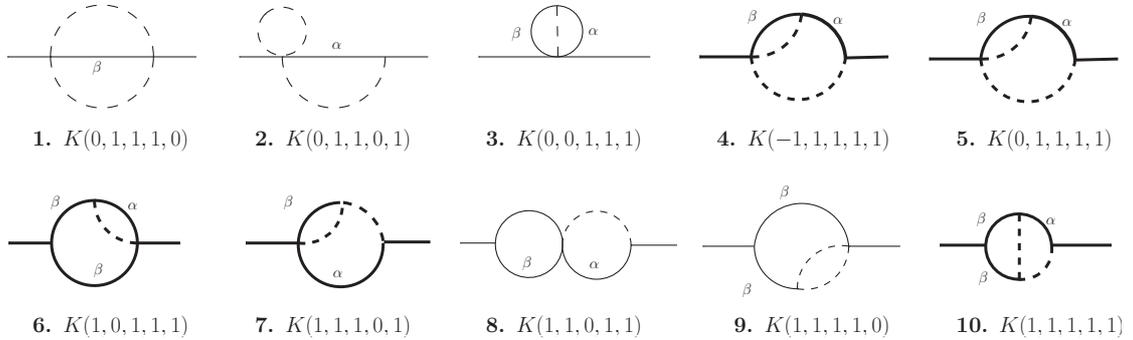}
\caption{Topologies for $f_2$.}
\label{f2}
\end{figure}

For the other self-energy in figure \ref{selfE}, which contains no branch cuts, one could in principle carry 
out the same calculation. The first step would be to reduce the main diagram to a sum of scalar integrals (figure \ref{f2}),
\noindent which gives a tensor decomposition into scalar integrals of the form
\begin{align}
f_2=  \slashed{q} \frac{M_\alpha}{M_\beta} \Bigg[  & \frac{1}{2 q^2} \Big( K( 0,1,1,1,0) - K(0,1,1,0,1) - K(0,0,1,1,1) + K(-1,1,1,1,1)\Big)  
\nonumber \\
& + \left[  \frac{2 M_\beta^2+ M_\alpha^2 }{2 q^2}  -\frac{1}{2}  \right] K(0,1,1,1,1) + 
\left[  \frac{1}{2} - \frac{M_\beta^2}{q^2} \right] K(1,0,1,1,1)   - \frac{1}{2} K(1,1,0,1,1) \nonumber \\
&- \frac{M_\beta^2}{q^2} K(1,1,1,0,1)   +  \frac{M_\beta^2}{q^2} K( 1,1,1,1,0)  + 
\left( - \frac{M_\beta^2}{2} +\frac{M_\alpha^2  M_\beta^2}{ q^2}   \right) K(1,1,1,1,1)  \Bigg] ,
\label{f2tensordecomp}
\end{align}
where now
\begin{align}
& K(\nu_1,\nu_2,\nu_3,\nu_4,\nu_5)  \nonumber \\
&= \int \frac{d^d k_1}{(2\pi)^d}\int \frac{d^d k_2}{(2\pi)^d}
\frac{1}{ [k_1^2- M_\beta^2]^{\nu_1}[ k_2^2 - M_\beta^2 ]^{\nu_2}  \left[ (k_1+k_2)^2 \right]^{\nu_3}  
\left[ (k_1-q)^2 \right]^{\nu_4}\left[ (k_2+q)^2 - M_\alpha^2 \right]^{\nu_5}}.
\end{align}

Continuing with this evaluation nevertheless requires very substantial further work and we restrict ourselves
here to some observations about the sterile neutrino mass dependence in the case of a large hierarchy.
By power counting in (\ref{f2tensordecomp}) and the diagrams in figure (\ref{f2}), we can see that only 
some of the diagrams (shown in bold) have the potential to give a large $\xi$ behaviour of the same order 
as in  (\ref{F1hierarchy}). 
The way to see this is to note that since none of them contains branch cuts, each one is analytic in $q^2$ and, 
as explained in \cite{Davydychev} (see eq.(2.6) therein), can be Taylor expanded before performing any momentum integrations. 
Each Taylor expansion can be written as a sum of 3-mass vacuum diagrams, each of which is given in terms of 
hypergeometric functions, as in (4.3) of \cite{Davydychev}. One can then power count the masses $M_\alpha$ and $M_\beta$ 
in these formulae for each of the diagrams 4,5,6,7 and 10. Carrying out this procedure, we find that in both limits 
$M_\alpha \gg M_\beta$ and $M_\alpha \ll M_\beta$, the asymptotic behaviour is never stronger than in  (\ref{F1hierarchy}) . 

An open question is then whether these contributions to $f_2(q,M_\alpha,M_\beta)$ in (\ref{4decomp}) can eventually 
conspire to exactly cancel the leading order behaviour of $f_1(q,M_\alpha,M_\beta)$ shown in (\ref{F1hierarchy}), or
whether the overall mass hierarchy dependence from diagram (4) remains of the form 
$I_{[\alpha\beta]}^{(4)}\simeq O(\xi^2 \log\xi^2)$ we have found for $f_1(q,M_\alpha,M_\beta)$.

\subsection{Yukawa couplings}

\begin{figure}[h]
\centering
\includegraphics[scale=0.35]{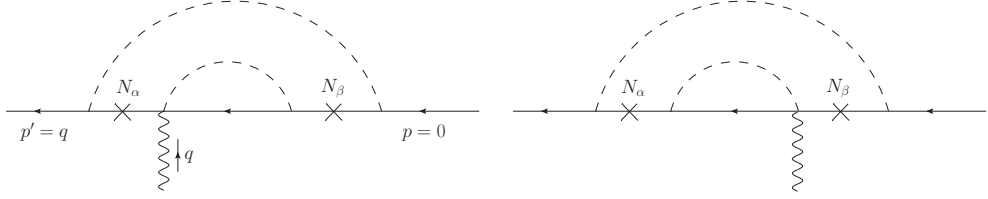}
\caption{Contributions from Yukawa insertions. }
\label{Yukh}
\end{figure}

The contribution from the $h$ coupling to the Yukawa interaction in (\ref{hinsertion}) is proportional to $(n-4)$, 
and so only contributes in diagrams which produce poles $1/(n-4)$. The vertex correction diagram is UV finite, 
and so the only source of UV divergences comes from the propagator correction amplitude, via its one-loop sub-graph. 
In fact, for the purpose of calculating the antisymmetric quantity $I_{[\alpha \beta]}$, only two 
propagator correction diagrams with graviton insertions contribute. They are shown in figure \ref{Yukh}.

Notice that we do not calculate the contributions from Yukawa insertions on the outermost vertices. 
The reason for this is that an insertion on the vertex of the outgoing lepton carries momentum $q$ in
and straight back out again, so that the diagram has no $q$ dependence. Similarly, the insertion of $h$ 
on the vertex of the incoming lepton gives only a symmetric contribution to $I_{\alpha \beta}$. 
Moreover, in this case, the divergences in the sub-diagram are removed by renormalization counterterms in the 
corresponding one-loop diagrams.
Hence, the only diagrams of interest are the two shown above. They give a contribution
\begin{align}
I_{[\alpha \beta]} &= - \frac{1}{48}\frac{1+x}{(1-x)^4} \left[ (1-x)(1 + 10 x + x^2) + 6 x (1+x ) \ln x\right] \nonumber \\
&\simeq - \frac{1}{48} + O\left(\frac{\ln x}{x}\right) \ ,
\end{align} 
for large $x$.  

\subsection{Higgs couplings}

If we consider a conformally-coupled Higgs Lagrangian with $\zeta = 1/6$, then from (\ref{hinsertion}) 
the only $h$ couplings to the Higgs propagators are proportional to the mass $m_H^2$. 
These contributions will therefore be highly suppressed relative to the insertions on the sterile neutrino
propagators. For a non-conformal Higgs coupling, (\ref{hinsertion}) shows there are further $h$ insertions 
proportional to $\left(\zeta - \frac{1}{6}\right) \partial^2 h$. These have the potential to contribute in a similar
way to the Yukawa terms discussed above.

\section{Leptogenesis and Baryogenesis}\label{leptogenesis}

It is now clear how the mechanism of radiatively-induced gravitational leptogenesis emerges in this see-saw model. 
First, matter and antimatter propagate differently by virtue of the breaking of time-translational invariance by the 
gravitational background, together with CP violation from the phases of the complex Yukawa couplings.
Both of these are radiatively-induced features which arise at two loops, as discussed in section \ref{YFandRIGL}. 
This is manifest in the difference of the matter and antimatter self-energies, {\it i.e.} $\Sigma(x,x') \neq \Sigma^c(x,x')$.
This bias in the dynamics of matter and antimatter is reflected in the particular operator (\ref{Operator}), 
whose effective coupling constant depends both on the imaginary part of the Yukawa couplings and on the sterile neutrino
masses via the quantity $I_{[\alpha \beta]}$, which is determined by the two-loop self-energy diagrams for the light leptons. 
Naturally, these are the same diagrams (figures \ref{2loopagain} and \ref{2loops}) which lead to distinct matter
and antimatter propagation. 

\subsection{Lepton asymmetry in a radiation-dominated FRW background}

We now show how the lepton asymmetry arises in the simplest possible model, where the background
is a radiation-dominated FRW spacetime at temperature $T$. The coupling in (\ref{Operator}) then acts as an 
effective chemical potential, so that in the thermal background of the hot early Universe,
the equilibrium distributions of the light leptons and antileptons are different, giving rise to a net lepton number
as shown in (\ref{Operator}) - (\ref{RIGL}). We therefore find the lepton-to-photon ratio $Y_L$ in equilibrium is given by
\begin{equation}\label{YL}
Y_L  \simeq \frac{ \pi^2 \dot{R}}{2 \zeta (3) T} \sum_{\alpha,  \,  \beta, \,   j,i} \frac{\mbox{Im}\left[
    \lambda^\dagger_{\beta i} \lambda_{i\alpha} \lambda^\dagger_{\beta
      j} \lambda_{j \alpha} \right]}{18 M_\alpha M_\beta} I_{[\alpha\beta]}(\xi),
\end{equation}
where we have used the photon density  $n_\gamma = 2 \zeta(3)T^3/\pi^2$. 

We can trace the origin of each term in (\ref{YL}) back to the fundamental principles set out 
in the introduction. First, $\dot{R}$ arises due to the breaking of time-translation symmetry
by the background geometry, the factor $I_{[\alpha\beta]}(\xi)/M_\alpha M_\beta$ describes the dependence of the loops
on the sterile neutrino mass hierachy, while
$\mbox{Im}\left[\lambda^\dagger_{\beta i} \lambda_{i\alpha} \lambda^\dagger_{\beta  j} \lambda_{j \alpha} \right]$
 arises from the breaking of C and CP symmetry. 

At this point, we can also see why the diagrams containing charge-conserving sterile neutrino propagators do not contribute.
(See section \ref{flatpropagation}, especially expression (\ref{chargeconserving}) for the relevant self-energy diagram.)
Since in this case the contribution from the vertices is
$\left[\lambda^\dagger_{\beta i } \lambda_{i \alpha} \lambda^\dagger_{\alpha j} \lambda_{j \beta}\right]$,  
we find $n(L) - n(L^c)  \propto \sum_{\alpha,  \,  \beta}\mbox{Im}
\left[ (\lambda^\dagger \lambda )_{\beta \alpha} ( \lambda^\dagger \lambda)_{\alpha \beta}\right] J_{\, [\alpha\, \beta]}$, 
for some loop factor $J_{\alpha \beta}$, when we sum over all generations. However, $\mbox{Im}\left[ (\lambda^\dagger  \lambda )_{\beta \alpha}
( \lambda^\dagger \lambda)_{\alpha \beta}\right] = 0$, and so the total lepton asymmetry from these diagrams vanishes. 

Now, the Ricci scalar in a FRW universe dominated by matter with equation of state parameter $w$ satisfies
\begin{equation}\label{Ricci}
R = -(1-3w) \frac{\rho}{M_p^2}, \qquad \qquad  \dot R = \sqrt{3} (1-3w)(1+w) \frac{\rho^{3/2}}{M_p^3},
\end{equation}
where $M_p$ is the reduced Planck mass and the expression for $\dot R$ follows from the conservation and
Friedmann equations. For the classically conformal invariant case of radiation dominance, $w= 1/3$.
At the quantum level, however, the energy-momentum tensor has a trace anomaly and the factor $(1-3w)$
acquires a contribution from the beta functions characterising the particle content of the theory
\cite{Davoudiasl:2004gf}. Below, we take $1-3w \simeq 0.1$. Since for radiation, $\rho = \sigma T^4$ with 
$\sigma = \pi^2 g_*/30$, where $g_*$ counts the effective degrees of freedom, we find the time derivative of the 
curvature at temperature $T$ is
\begin{equation}\label{Riccidot}
\dot R = \sqrt{3} \sigma^{3/2} (1-3w)(1+w)\frac{T^6}{M_p^3} \ .
\end{equation}
Substituting back into (\ref{YL}) we find
\begin{equation}\label{YLT}
Y_L  \simeq \frac{\sqrt{3}\pi^2 \sigma^{3/2}(1-3w)(1+w)}{36\zeta(3)} \frac{T^5}{M_p^3}
\sum_{\alpha,  \,  \beta, \,   j,i} \frac{\mbox{Im}\left[
    \lambda^\dagger_{\beta i} \lambda_{i\alpha} \lambda^\dagger_{\beta j} \lambda_{j \alpha} 
    \right]}{M_\alpha M_\beta} I_{[\alpha\beta]}(\xi).
\end{equation}

At this point, we can return to the discussion of the validity of the effective Lagrangian in section \ref{EFTcurved}.
Using the expression (\ref{Ricci}) for the curvature, the weak gravitational field and low-energy conditions
can be re-expressed in terms of the temperature as
\begin{equation}
\frac{T^2}{M_1 M_p}  \lesssim  1, \qquad   \frac{T^3}{M_1^2 M_p} \lesssim 1 \ ,
\end{equation}
respectively, where we have taken the typical lepton energy as $E\sim T$ and $M_1$ as the sterile neutrino mass. 
These will clearly be satisfied in the region of interest, $T \sim M_1$ and $T \ll M_p$.
This means that the prediction (\ref{YLT}) calculated from the effective Lagrangian is valid provided
the temperature factor $\sim T^5/M_1^2 M_p^3$ in $Y_L$ is small, which is certainly the case observationally.

\subsection{Towards leptogenesis and baryogenesis}

Finally, we discuss briefly how this non-vanishing equilibrium lepton asymmetry may play 
a r\^ole in determining the baryon-to-photon ratio $\eta_B$ of the Universe. 
We assume a standard leptogenesis scenario in which a lepton asymmetry established at relatively high
temperature in a radiation-dominated FRW universe can subsequently be transformed into a baryon asymmetry 
by the conventional sphaleron mechanism \cite{Kuzmin:1985mm} (see \cite{Buchmuller} for a summary) when the
temperature has dropped below the usual sphaleron scale $\sim 10^{12}$ GeV. From the observed value
$\eta_B \simeq 6 \times 10^{-10}$, we infer the corresponding ratio
$Y_L \simeq 3 \times 10^{-8}$.\footnote{Here we have used the relation 
\begin{equation*}
Y_B = C_{sph} Y_{B-L} = \frac{C_{sph}}{1-C_{sph}} Y_L \ ,
\end{equation*}
with $C_{sph} = 28/79$ in the standard model, and included the standard factor 
$f = 2387/86$ to account for the production of photons from the leptogenesis scale to CMB formation
\cite{Pedestrians}. }

To get an initial orientation on the relevant orders of magnitude,
consider the scenario where the sterile neutrino masses satisfy $M_1  \ll M_2 \ll M_3$. 
If we assume the mass dependence found in (\ref{F1hierarchy}) holds for the complete contribution
from diagram (4), the asymmetry $Y_L$ in (\ref{YLT}) will be dominated by the largest hierarchy,
{\it i.e.}~$\xi = M_3/M_1$, with $I_{[13]} \simeq \xi^2 \log{\xi^2} /(4\pi)^4$.
It is also convenient to introduce the standard parameter $z = M_1/T$.
This gives our key result (\ref{YLT}) in the form
\begin{equation}\label{YLparameters}
Y_L \simeq   \alpha_{13}^2 \sin\delta\,\frac{\xi \ln{\xi^2}}{z^5}\left(\frac{M_1}{M_p} \right)^3  
\left[ \frac{\sqrt{3} \sigma^{3/2}\pi^2 (1-3w) (1+w) }{36 \zeta(3)(4\pi)^2} \right]  \ , 
\end{equation}
where $\alpha_{13} = \left| ( \lambda^\dagger \lambda )_{13}\right|/(4\pi)$ is the appropriate coupling constant, 
and  $\delta = \mbox{Arg}[ ( \lambda^\dagger \lambda )_{13}^2]$ quantifies the size of CP violation. 
In the numerical estimate below, we take $\alpha_{13} \simeq 0.8$ and $\sin\delta \simeq 1$.
Inserting these values, together with $(1-3w) = 0.1$, $g_* = 106.75$ and $M_p = 2.4\times 10^{18}$ GeV,
we obtain the following expression for the equilibrium lepton-to-photon ratio at temperature $T$:
\begin{equation}\label{YLnumbers}
Y_L \simeq  4.4 \times 10^{-3}\, \frac{\xi \ln{\xi^2}}{ z^5} \left(\frac{M_1}{M_p} \right)^3 \ .  
\end{equation}
\begin{figure}[h!]
\centering
\includegraphics[scale=1]{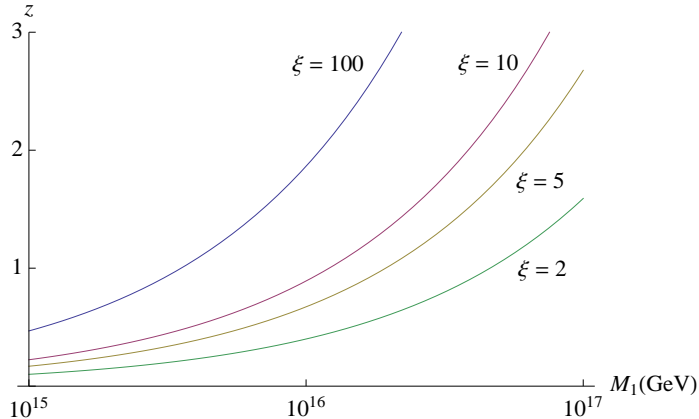}
\caption{The curves show the values of $z= M_1/T$ and $M_1$ which give the lepton-to-photon ratio 
$Y_L = 3\times 10^{-8}$ in an observationally important range, for various values of the 
sterile neutrino mass hierarchy parameter $\xi$. }
\label{parameters} 
\end{figure}

In figure \ref{parameters}, we plot contours with $Y_L = 3\times 10^{-8}$ in the $z,M_1$ plane for 
different values of the hierarchy parameter $\xi$. This shows that the gravitationally-induced lepton asymmetry
may indeed be of the order of magnitude necessary to play a r\^ole in determining the observed baryon-to-photon ratio, 
with sterile neutrino masses and temperatures in the GUT range $\sim 10^{16}$ GeV.

In the see-saw model, which we have used here to illustrate the radiatively-induced gravitational mechanism, 
the sterile neutrino masses and hierarchy determine the light neutrino mass spectrum and
these values correspond to neutrino masses of the order of a few $10^{-3}$ eV, putting them in the lower end of the range 
allowed by solar and atmospheric neutrino data \cite{Buchmuller}.
Notice, however, that the fundamental gravitational effect, which gives rise to the chemical potential $\mu \sim \dot{R}/M_1^2$,
is in general actually favoured by lower sterile neutrino masses. The relatively high mass parameters discussed above are
being driven by the assumption that the relevant temperatures at which the lepton asymmetry freezes out are $z \gtrsim 1$.

If we consider the model in the context of the exit from inflation, then, as discussed in \cite{Davoudiasl:2004gf}, 
this decoupling temperature must satisfy $T \lesssim T_{RH} \lesssim V^{1/4}$,
where $T_{RH}$ is the reheat temperature and $V^{1/4}$ gives the inflationary scale. 
We would therefore require an inflationary scale of the order of the GUT scale, 
with the lepton asymmetry freezing out at high temperatures near the beginning of radiation dominance.
This corresponds to a value of the tensor-to-scalar ratio $r$ (characterising the strength of the gravitational
waves produced by inflation) to be close to the current upper bound $r_{max} = 0.07$ \cite{Array:2015xqh}.

The simplest scenario, implicit in the discussion above, is that the lepton asymmetry freezes out at its equilibrium value
when the reaction rates for the lepton-number violating reactions, which maintain the leptons and antileptons in thermal 
equilibrium with the asymmetry (\ref{YLT}), fall below the Hubble expansion parameter.
Of course, this is only a simple first approximation.
In general, finding the physically realised lepton asymmetry in this model will depend on a detailed dynamical analysis 
of all the simultaneous reactions and decays taking place around the critical scale $T\simeq M_1$. This would 
involve a full treatment of the coupled Boltzmann equations, taking into account initial abundances, inverse decays,
and $\Delta L=1$ and $\Delta L=2$ scattering rates. This is essential to determine how closely the actual lepton
asymmetry is tracking its equilibrium value at the point of freeze-out.
We also need to include near-resonant production of the sterile neutrinos and their out-of-equilibrium decays,
since in this model the original leptogenesis mechanism described in section \ref{YFandRIGL} is simultaneously active.

This complete analysis of the kinetic theory of the model in its cosmological setting is currently under investigation
and will be presented elsewhere. Here, our motivation has been simply to demonstrate that the new mechanism of
radiatively-induced gravitational leptogenesis can produce asymmetries of the required order of magnitude to account
for, or certainly play an important r\^ole in determining, the observed baryon-to-photon ratio in the Universe.

\section{Outlook}

In this paper, we have presented a new mechanism for leptogenesis and baryogenesis in which the matter-antimatter 
asymmetry is generated by gravitational couplings induced by quantum loop effects in curved spacetime.
In this mechanism, the Sakharov conditions are realised as follows. The first occurs in the usual way for leptogenesis
through lepton number violating reactions mediated by heavy BSM particles. C and CP violation can arise from
complex phases in the coupling of the light leptons to these heavy states. The time dependence necessary, 
in the spirit of the third Sakharov condition, for generating a matter-antimatter asymmetry arises not by the traditional 
mechanism of out-of-equilibrium decays of the BSM particles, but through the direct CP and SEP-violating coupling of
quantum loops to the time-dependent gravitational field in the expanding Universe. In this scenario, the heavy
BSM states contribute only as virtual particles to the self-energy cloud screening the light leptons. The lepton
number asymmetry is then transferred to a baryon asymmetry via the usual sphaleron mechanism.

Specifically, we showed in the context of the see-saw model how the virtual sterile neutrinos in the two-loop
self-energy contributions to the light lepton propagators allow the leptons to become sensitive to the time-dependent
dynamics of the gravitational background and to the C and CP violation in the BSM Lagrangian. This curvature coupling
breaks the strong equivalence principle and allows particles to propagate differently; the sensitivity to CP violation then 
allows a distinction between the dispersion relations for leptons and antileptons. This effect induces an effective chemical
potential, which modifies the equilibrium distributions of leptons and antileptons and allows a lepton number
asymmetry to be maintained in the thermal quasi-equilibrium characterising the early radiation-dominated Universe.
Remarkably for an intrinsically QFT effect in curved spacetime,
this effect is sufficiently strong to play a r\^ole in determining the observed baryon-to-photon ratio. This is because, 
although the loop effects we have calculated are necessarily very small, they are the leading symmetry-breaking 
contribution to a quantity which would otherwise be zero by translation
invariance.\footnote{Another well-known example where curved spacetime QFT effects are important in cosmology is the study 
of quantum fluctuations in the inflationary phase of the Universe, which can be indirectly probed by CMB measurements.}. 

This work suggests many areas for further investigation. The most immediate is to embed the mechanism into a detailed study 
of lepton number generation in the early radiation-dominated Universe, taking into account the interplay between the conventional
out-of-equilibrium decays of on-shell sterile neutrinos and our new mechanism of radiatively-induced gravitational leptogenesis.
This will involve a full analysis of the temperature-dependent, kinetic aspects of the evolution, including coupled Boltzmann
equations, reaction rates and freeze-out temperatures as the Universe cools. An interesting question is to determine whether one 
mechanism dominates over the other for particular parameter ranges in the fundamental BSM theory.

It is important to emphasise, however, that the leptogenesis mechanism we are proposing is far more general than its
realisation in the particular see-saw model presented in this paper. It will arise notably in generic BSM theories
exhibiting C and CP violation at a high energy scale. The main condition is that since this is a gravitational leptogenesis
mechanism, the matter-antimatter asymmetry must be generated at a sufficiently early time  
that the curvature of the Universe is still strong enough to produce the observed baryon-to-photon ratio. 
Since this scale may also be characteristic of the temperatures at the end
of inflation, it will be interesting to look at scenarios where our mechanism is embedded into inflationary 
models.\footnote{See, for example, \cite{Alexander:2004us, Berera} for models of leptogenesis in inflation.}
In principle, it can also be applied directly to the generation of a baryon asymmetry through a radiatively-generated gravitational
coupling to the baryon number current.

On the theory side, the central idea underlying our mechanism is that in the presence of a time-varying gravitational field,
matter and antimatter in a C and CP-violating theory propagate differently at loop level. In this paper, we translated this 
fundamental observation into a mechanism for leptogenesis by first using the effective Lagrangian formalism to identify the 
relevant operator (\ref{Operator}), then interpreting its coupling as a chemical potential which changes the dispersion relations
and induces a difference in the equilibrium distributions for matter and antimatter. We showed that this operator arises
naturally through radiative corrections in curved spacetime, without the need to appeal to an as yet unknown theory of 
quantum gravity. While this approach is justified in the early Universe where the traditional quasi-equilibrium 
approach to kinetic theory is a good approximation, from a theoretical perspective we would like to develop a more
fundamental analysis. An ideal strategy would be to describe the lepton (baryon) asymmetry {\it directly} 
from the self-energies $\Sigma(x,x')$, which should be treated within a real-time, non-equilibrium, curved spacetime
framework to calculate the time evolution of the lepton (baryon) number. This would provide a theoretically rigorous,
real-time formulation of radiatively-induced gravitational leptogenesis.

\subsection*{Acknowledgements}

One of us (J.I.~McD) would like to thank B.~Garbrecht and V.~A.~Smirnov for helpful correspondence and advice.
We are grateful to G.~Aarts, D. De Boni and T.~Hollowood for many useful conversations.
This research is supported in part by STFC grants ST/K502376/1 and ST/L000369/1.


\appendix 
\section{Techniques for evaluating the self-energy diagrams}

In this appendix, we demonstrate some of the techniques used for evaluating the Feynman diagrams in section \ref{Results}. 
For instance, in the case of diagram (1), after inserting the sub-diagram into the remaining momentum integral, we have
\begin{align}
& f^{(1)}= \frac{M_\alpha}{2 (4\pi)^2} \int \frac{d^d k }{(2\pi)^d} \frac{ \left[ (\slashed{k} 
+ \slashed{q}) \slashed{k} + M_\alpha^2 \right] \slashed{k}}{k^2 [(k+q)^2 -M_\alpha^2][k^2 - M_\alpha^2][ k^2 - M_\beta^2]}
\left( 1 - \frac{1}{2} \ln \left( \frac{-k^2}{\mu^2}\right)\right).
\end{align}
The constant term from $\Sigma(k)$ can be evaluated in the usual way, by manipulation of the numerator 
and introduction of Feynman parameters.  The logarithmic term, however, is more subtle, but can be dealt 
with by noting we can write
\begin{equation}\label{slog}
 \ln \left( \frac{-k^2}{\mu^2} \right) = \lim_{s \rightarrow 0}\frac{d}{ds} \left[ \frac{(- \mu^2)^s}{\left[ k^2 \right]^s}\right]. 
\end{equation}
This trick can be employed for any other logarithmic term generated by a sub-diagram. 
This reduces the calculation to the evaluation of a Feynman integral containing a denominator factor 
raised to an arbitrary power $s$, \textit{viz.}
\begin{equation}
I(s) \equiv \frac{M_\alpha}{2 (4\pi)^2} \int \frac{d^d k }{(2\pi)^d} \frac{ \left[ (\slashed{k} 
+ \slashed{q}) \slashed{k} + M_\alpha^2 \right] \slashed{k}}{k^2 [(k+q)^2 -M_\alpha^2][k^2 - M_\alpha^2][ k^2 - M_\beta^2][k^2]^s} ,
\end{equation}
which can then be differentiated with respect to $s$ after performing the momentum integration.
 First, we note that the numerator can be rewritten as
\begin{equation}
 (\slashed{k} + \slashed{q}) \slashed{k} + M_\alpha^2  = (\slashed{k} + \slashed{q})(k^2 -M_\alpha^2) 
+ (2\slashed{k} + \slashed{q})M_\alpha^2,
\end{equation}
allowing us to cancel the numerator factor $(k^2 - M_\alpha^2)$ against a denominator, so that $I(s)$ 
splits into two simpler integrals. For instance, the second term gives a contribution
\begin{equation}
I (s)=\int \frac{d^d k}{(2\pi)^d} \frac{(2  \slashed{k} + \slashed{q}) (-\mu^2)^{s}}{[ k^2 ]
[(k+q)^2 - M_\alpha^2][k^2 - M_\alpha^2] [k^2-M_\beta^2] [k^2]^s},
\end{equation}
from which we need only the $O(s)$ term in accordance with (\ref{slog}). After introducing Feynman parameters, we get
\begin{equation}
I (s)=  \frac{\slashed{q}}{(4\pi)^2} \left( \frac{\mu^2}{M_\alpha^2}\right)^s \frac{s }{ M_\alpha^2} \int_0^{1} dy \int_0^{1-y} dw  
\int^{1-y-w }_0 \! \! \! \! dz  \frac{  \, w^{s-1} (y-1)  }{\left[ Q y (1-y) + y + z b \right]^{1+s}},
\end{equation}
where
\begin{equation}
Q = - q^2/M_\alpha^2, \qquad b = M_\beta^2/M_\alpha^2.
\end{equation}
Performing the $z$ and then $w$ integrals gives an answer in terms of incomplete Beta functions:
\begin{align}
I(s)= \frac{\slashed{q}}{ (4\pi)^2 } \left( \frac{\mu^2}{M_\alpha^2}\right)^s \frac{1}{ M_\beta^2} \int_0^{1} dy     \,  (1-y) 
& \Big\lbrace\frac{1}{b^s}\,B\left[ \frac{b (1-y)}{Q y (1-y) + y + z (1-y-w)} ,s,1-s\right] \nonumber \\
& - \frac{1}{s}\frac{(1-y)^s}{\left[ Q y (1-y) + y \right]^{s}}  \Big\rbrace.
\end{align}
Using the Taylor series for the incomplete beta function, we find the $s$ expansion gives
\begin{equation}
\mbox{B}[x,s,1-s] = \frac{1}{s} + \ln(x) + s \left[  \frac{1}{2} \ln^2 (x) + \mbox{Li}_2(x)\right] +O(s^2).
\end{equation}
Notice that poles in $s$ cancel between the Beta function and the second term in $\left\{... \right\}$. 
Taking the term linear in $s$ gives
\begin{align}
\lim_{s \rightarrow 0}\frac{d I(s)}{ds} = \frac{\slashed{q}}{ (4\pi)^2 }\frac{1}{ M_\beta^2} \int_0^{1} dy     \,  (1-y) 
\Bigg\lbrace & \frac{1}{2} \ln^2 \left( \frac{\mu^2}{M_\beta^2}\right) + \ln\left( \frac{(1-y)b}{A}\right) \ln (L_\beta)  \nonumber \\
& + \frac{1}{2} \ln^2\left( \frac{(1-y)b}{A}\right) + \mbox{Li}_2 \left( \frac{(1-y)b}{A} \right) \nonumber \\
& - \frac{1}{2}\ln^2 \left(\frac{(1-y)}{ Q y (1-y) +y} \right)\Bigg\rbrace,
\end{align}
where
\begin{equation}
A = Q y (1-y) + y + (1-y)b. 
\end{equation}
A full expression for the remaining $y$ integration is too lengthy to write down here. 
However, each of the terms in the integrand is analytic in $q$ for all $y$ in the integration range, 
provided $-q^2 \ll M_\alpha^2, M_\beta^2$, and so the integrand can be analytically expanded in powers 
of $q^2/M_\alpha^2$ {\it prior} to performing the $y$ integral. This is a reflection of the fact this diagram 
contains no branch cuts, {\it i.e.}, no zero-mass thresholds, and thus has no $\log \left(-q^2 / M_\alpha^2\right)$ terms. 
Expanding in $-q^2/M_\alpha^2$, and then performing the $y$ integral, we find an answer of the form
\begin{equation}
\lim_{s \rightarrow 0} \frac{d I(s)}{ds}
=  \frac{\slashed{q} }{(4 \pi)^2} \frac{1}{M_\beta^2} \sum_n C_n \left( \frac{-q^2}{M_\alpha^2}\right)^n,
\end{equation}
where for our purposes the relevant coefficient is
\begin{equation}
C_1 =  -\frac{(b ((b-6) b+3)+6 b \ln (b)+2) \ln\left(\frac{\mu^2}{M_\alpha^2} \right)-3 b\ln^2 (b)+3 (b-1)^2}{6
   (b-1)^4 }  -\frac{2  \ln\left(\frac{\mu^2}{M_\alpha^2} \right)+3}{6 }.
\end{equation}
We can repeat this exercise for the other integrals contributing to diagram (1) to arrive at the
answer quoted in the text in (\ref{I1}).

Similar techniques are used to evaluate the other self-energy diagrams, with the results
quoted in section \ref{Results}.


\end{document}